\documentclass[article]{JHEP}
\usepackage{amsmath,epsfig}
\usepackage{amssymb,amsfonts}
\usepackage{epsfig}
\relax
\renewcommand{\theequation}{\arabic{section}.\arabic{equation}}
\def\be{\begin{equation}}
\def\ee{\end{equation}}

\newcommand\fverb{\setbox\pippobox=\hbox\bgroup\verb}
\newcommand\fverbdo{\egroup\medskip\noindent%
                        \fbox{\unhbox\pippobox}\ }
\newcommand\fverbit{\egroup\item[\fbox{\unhbox\pippobox}]}
\newcommand{\nn}{\nonumber}

\newbox\pippobox

\newcommand{\de}{\partial}

\renewcommand{\a}{\alpha}
\renewcommand{\b}{\beta}

\renewcommand{\d}{\delta}

\def\hre#1#2{\href{http://arxiv.org/abs/#1/#2}{[ArXiv:#1/#2]}}
\def\hrep#1#2{\href{http://arxiv.org/pdf/#1/#2v1}{[ArXiv:#1/#2]}}

\def\dalemb#1#2{{\vbox{\hrule height .#2pt
        \hbox{\vrule width.#2pt height#1pt \kern#1pt
                \vrule width.#2pt}
        \hrule height.#2pt}}}

\def\lsim{\mathrel{\rlap{\lower4pt\hbox{\hskip1pt$\sim$}}
    \raise1pt\hbox{$<$}}}                
\def\gsim{\mathrel{\rlap{\lower4pt\hbox{\hskip1pt$\sim$}}
    \raise1pt\hbox{$>$}}}                

\def\slash{\raise.15ex\hbox{$/$}\kern-.57em}
\def\slashA{\raise.15ex\hbox{$/$}\kern-.70em A}

\newcommand{\Real}{\mathbb{R}}

\def\a{\alpha}
\def\b{\beta}
\def\d{\delta}
\def\e{\epsilon}

\def\m{\mu}
\def\n{\nu}

\newcommand{\beq}{\begin{equation}}
\newcommand{\eeq}{\end{equation}}
\newcommand{\bea}{\begin{eqnarray}}
\newcommand{\eea}{\end{eqnarray}}

\title{On massless 4D Gravitons from Asymptotically $AdS_5$ Space-times}
\author{
E. Kiritsis$^{1,2}$, F. Nitti$^{1}$\\
$^1$CPHT, Ecole Polytechnique, CNRS,
 91128, Palaiseau, France\\
 ( UMR du CNRS 7644).\\
~\\
$^2$Department of Physics, University of Crete\\
71003 Heraklion, Greece\\
~\\}

\preprint{\hepth{0611344} \\ CPHT-RR019.0503 }

\abstract{We investigate the conditions for obtaining  four-dimensional
massless spin-2 states  in the spectrum of fluctuations 
around an asymptotically $AdS_5$  solution of
 Einstein-Dilaton gravity. We find it is only possible to
have  normalizable massless spin-2 modes if the space-time
terminates at some IR  point in the extra dimension, far from the UV
AdS boundary,
 and if suitable boundary conditions are imposed at the ``end of space.''
 In some of these  cases the
4D spectrum consists only of a massless spin-2 graviton,  
with no additional massless or light scalar
 or vector modes. These spin-2 modes have a profile 
wave-function peaked  in the interior of the 5D bulk space-time.    
Under the holographic duality, they 
may be sometimes interpreted as  arising purely from the IR dynamics of a
 strongly coupled QFT living on the AdS boundary.}

\begin{document}

\maketitle 

\section{Introduction and Summary}

Gravity remains to date the least understood of all known fundamental interactions.
Recent cosmological data, imply the presence of
energy sources in the universe resembling a vacuum energy.
This is hard to explain in theories of fundamental physics that successfully
incorporate gravity, notably string theory.

String theory provides a perturbatively well defined quantum
theory of gravity at energies well below the ten-dimensional Planck scale,
but seems to fail to naturally explain today's acceleration of the universe.
Although the theory has been successful in resolving several types
of space-time singularities, ubiquitous singularities
like the ones appearing inside black holes or
at the beginning of the universe have resisted resolution.
It is quite plausible that they cannot be resolved with a perturbative
treatment of the theory.
This is in contrast with the microscopic understanding of black
hole entropy in string theory, that was essentially successful.

The resolution of black-hole and cosmological singularities
should be correlated with the behavior of the theory at high energy.
String theory is strongly coupled at energies approaching the Planck scale, and strong coupling
 dualities are seemingly unable to help in this direction.
Closed string theory is a theory with an effective cutoff at the string scale $M_s$
 typically much smaller than the Planck scale $M_P$ in the perturbative regime.
 It is therefore not a surprise that the perturbative theory breaks down
at energies much higher than the effective cutoff.

The successful counting of black-hole microstates \cite{sv}
led to the AdS/CFT correspondence \cite{malda} and its generalizations.
This development gave a concrete realization to the correspondence between
large-N gauge theories and string theories as advocated
much earlier, \cite{hooft}.
An interesting byproduct of this correspondence is the use of gauge-theory dynamics, in order to
 understand non-perturbative and/or high energy issues in string theory on spaces with boundary.
 The old matrix model approach  could be reinterpreted
 in this new light as a two-dimensional bulk-boundary correspondence, \cite{mv}.
 One could go further and postulate that observable gravity is described by a
 four-dimensional large-N (gauge) theory, unifying in such a fashion
 gravity with the other interactions in four dimensions, \cite{gg1,gg2}.

 A glimpse of the five-dimensional gravitational setup, associated to
four-dimensional conformally invariant gauge theories is provided
 by the Randall-Sundrum geometry, \cite{RS}.
 The geometry is locally AdS$_5$, but the boundary of the space has been
 removed by placing a gravitating 4-brane in AdS$_5$
supplemented by a Z$_2$ orbifold identification with a fixed point at the brane position.
The dual gauge theory is a cutoff version of the
${\cal N}$=4 SYM theory, although at high energies
this identification is only qualitative, \cite{gg2}.
On the gravitational side the presence of the
UV cutoff allows for a  four-dimensional
 massless graviton, on top of the continuum
of the five dimensional KK gravitons.
From the four-dimensional gauge theory point of view,
the massless graviton is fundamental, and cannot be described by
the gauge theory alone.
This is consistent with the intuition that bulk particles with
wave-functions peaked at the UV brane are fundamental in the dual gauge theory,
while particles with wave-functions peaked in the IR are composite bound-states, \cite{apr}.

In this paper we will systematically investigate the question:
can a massless four-dimensional
graviton appear in five-dimensional warped geometries that
are asymptotically AdS?
This is qualitatively distinct from the RS setup, since here
the space has a UV boundary with no cut-off,
and the standard RS graviton wave-function
is non-normalizable. In this sense we are searching for
massless composite gravitons, since necessarily
the graviton wave-function must have an important
support in the IR section of the geometry.

One of the main motivations for this exercise is that
we want to imagine that the five-dimensional gravitational theory
is a dual version of a four-dimensional large-N theory,
and the graviton is a composite of gluons.
In this context, the generic relevant bulk fields associated
to a gauge theory are  the metric, dual to the stress tensor
of the YM theory;
a scalar $\Phi$ (the dilaton) associated to the operator
$Tr[F_{\m\n}F^{\m\n}]$; and a pseudoscalar $A$ (the axion),
associated to the operator
$Tr[F\wedge F]$. As the axion is understood not to affect
the vacuum structure of the gauge
theory at large N, because it couples to instantons, we will
neglect it for most of the paper.

The question whether one can find a ``composite'' 4D graviton
 using arguments from holography was first investigated in
\cite{gherghetta}, where a model was proposed in which the metric
 is the only bulk field, and the space-time is
a slice of $AdS_5$ like  in RSI. Bulk and boundary mass terms for
 the five dimensional metric fluctuations were
then added to the Einstein-Hilbert action at quadratic level to
 modify the profile of  the graviton zero-mode
in the extra dimensions, and it was found that for a particular
 choice of parameters there is indeed an IR-peaked zero-mode.
In our work, we consider more general
5D geometries, characterized by a non trivial dilaton background, i.e.
we perform a systematic analysis of the question whether 4D
massless graviton modes can appear in 5D Einstein-Dilaton theory
(as suggested by a dual gauge theory picture).
 We avoid  some of the drawbacks of \cite{gherghetta} by
dealing with a  generally covariant bulk theory: rather than modifying
 the graviton dynamics by hand with mass terms, we modify the
background geometry by considering a nontrivial scalar field in the bulk.
   Adding a dilaton is  in some sense  a minimal modification
of the bulk theory, but at the same time it is sufficient
    to carry on a complete  study  of the 5D geometries that
have 4D flat sections, due to the freedom in choosing the dilaton potential \cite{freedman}.
As we will later argue, our results are general and valid for any other set of bulk fields and actions.

The method we use to analyze the 4D spectrum parallels the
 analysis of the spectrum of 4D modes in brane-world models  with a bulk scalar
\cite{CEHS,Tanaka:2000er,Csaki:2000zn,peloso}, and it follows
 the procedure used in cosmological perturbation theory (see \cite{mukhanov}
for a review): we look at the linearized equations for classical
 fluctuations of the bulk fields (metric and dilaton, in our case),
 classify them according to their 4-momentum and representation
of the 4D Lorentz group, and look for normalizable solution of the
 corresponding linearized field equations.
In this context, various authors have computed the low energy
``glueball'' spectra of gauge theories which are dual to deformations of
 the archetypal $AdS_5\times S^5$ background, and that develop a
mass gap and exhibit confinement in the IR
  \cite{witten,Csaki:1998qr,Constable:1999gb,Brower:1999nj,Brower:2000rp,Apreda:2003sy}.
 In fact, our philosophy and our
   techniques are  very similar to the kind of constructions generally
referred to $AdS$/QCD \cite{Erlich:2005qh,DaRold:2005zs}
   (see \cite{csaki-reece} for an up-to-date list of references):
there, one tries to reproduce the known low energy features
   of QCD by engineering a specific ``phenomenological''  5D model.
In our paper we are not interested in QCD, but the method
    we follow is similar: we try to study which kinds of
 5D setups have a field theory dual that contains a massless
    spin-2 ``glueball'' in the low-energy spectrum. Since the
backgrounds we study are  asymptotically $AdS_5$, the dual theory
    will be asymptotically conformal (but not asymptotically free)
 in the UV, although we are not going to analyze  its detailed nature.

We also analyze the potential presence of other  massless fields,
like scalars and vector bosons
that may arise from the universal sector of the bulk theory.
This is important, as their presence
can make the gravitational theory clash with data.

The basic result of our paper can be summarized as follows.
 Consider an  asymptotically $AdS_5$ space-time with  metric of the form:
$$
ds^2 = a^2(y) \left(dy^2 + \eta_{\mu\nu}dx^\mu dx^\nu\right)
$$
This is generic, since any 5D warped space-time with 4D Lorentz
invariance can be brought to this form. What we show is:
\begin{itemize}
 \item \emph{There can be no massless 4D graviton modes unless
the space-time ``ends'' at a finite value of the extra coordinate, $y=y_0$.}
\end{itemize}
The ``end of space'' at $y_0$ can correspond  either to an IR
 brane, (analogous to the Planck brane of RSI,  as
it was the case in \cite{gherghetta}), or to a naked singularity.
In the latter case we  show that:
\begin{itemize}
\item \emph{It is possible in some cases to have  a massless
 spin-2 as the \emph{only}  propagating 4D massless
degree of freedom with no extra  scalars or vectors.}
\end{itemize}

 We believe that the singular case, although at first sight
less appealing,  may ultimately be more interesting
  than having a  boundary at some arbitrary cut-off point. If we want
to provide a mechanism for ending space-time at a
   finite value of the extra coordinate, it is preferable that this
    results automatically  from the dynamics,  rather than from
 a boundary put by hand at an arbitrary location.
     Unfortunately, we will see that although this is true for
the background (the location and type of singularity
     are purely determined by the bulk potential plus UV initial
 conditions), the aspect of the IR dynamics in which we
     are more interested, i.e. the spectrum of light modes, depends
in a crucial way on additional information that we must
     provide at the singularity, in the form of boundary conditions
for the fluctuations,  and cannot be deduced from bulk or
      UV-boundary physics. This ultimately makes our search for an
 emergent massless graviton only partially successful, since any
      attempt at resolving the singularity must reproduce those boundary
 conditions for which massless gravitons are in the spectrum.
       This makes our results non-universal, and dependent on the details
 of the dynamics of the singularity.

 It should be stressed at this point that gauge theories that confine in the IR
 and have a five-dimensional bulk dual at all scales are generically expected to
  have a semiclassical IR singularity that is resolved in the
 full string theory, \cite{GK}.

We should mention that  there is a theorem that forbids massless composite
gravitons in gauge
theories, with specific assumptions including relativistic invariance and
the existence of well defined particle states, \cite{ww}. There are known
exceptions,
like the gravitons in AdS$_5$, which violate one or most assumptions.
We will discuss this issue and develop its ramifications in the last section.

Our paper is organized as follows. In Section 2 we
present our general Einstein-Dilaton model.
 In Section 3 we perform the linear fluctuation analysis
of the various bulk modes, classified according to
 their 4D mass and Lorentz tensor properties. In Section 4 we tackle
the question of the normalizability of
 modes, and we systematically investigate   what kind of models have
the desired normalizable zero-mode  spectrum.
 In Section 5 we discuss the issues of boundary conditions for singular space-times.
  In Section 6 we present a concrete example, in which we derive
explicitly the effective four-dimensional
  coupling of the massless spin-2 zero-modes.
In Section 7 we present some ideas on how to circumvent some of
  the difficulties of this setup, in particular how
to evade the ``no-go'' argument exposed above. Some of the
   technical details are left to the appendix.

\section{Five-Dimensional  Dilaton-Gravity}

A generic gauge theory in four dimensions has three gauge-invariant operators of lowest dimension:
 the stress tensor, dual to a metric,
$Tr[F^2]$ dual to a scalar (the dilaton), and $tr[F\wedge F]$ dual to a  pseudoscalar, the axion.
As the axion couples to instantons and these are negligible at large N we will not consider it further.

Driven by these considerations we consider 5D gravity plus a
scalar field with a generic potential\footnote{with space-time
coordinates $x^A\equiv (x^\mu, y)$.
The metric has signature $(-++++)$;  capital Latin indexes $A, B, \ldots$ are 5D indexes,
 taking values $\mu, y$; Greek indexes
$\mu,\nu,\ldots$ are 4D indexes ranging from $0\ldots3$. A primes denote derivative w.r.t. $y$.},
\beq\label{action}
S = {1\over 2k_5^2}\int d^5 x \sqrt{-g} \Big(R - \de_A \Phi \de^A \Phi - V(\Phi)\Big),
\eeq
and background solutions of the form
\beq\label{b0}
ds^2 = a(y)^2\left(dy^2 + \eta_{\mu\nu}dx^\mu dx^\nu\right), \qquad \Phi(y,x^\mu)=\Phi_0(y)
\eeq

For later reference, we define the functions $B(y)$,  $z(y)$ as:
\beq\label{zed}
B(y)=-{3\over 2}\log a(y), \qquad  z(y) = {\Phi_0' \over (a'/a)}.
\eeq
Einstein's equations take the form:
\bea
&&\Phi_0'^2 = -3 \left({a''\over a} - 2{a'^2 \over a^2}\right) = 2B'' + {4\over 3}B'^2, \label{b4}\\
&& a^2 V(\Phi_0) = -3\left({a''\over a} + 2{a'^2 \over a^2}\right) =  2B'' - 4B'^2 \label{b5}.\eea
and they imply the Dilaton field equation, as usual.
The 5D curvature scalar\footnote{this can be obtained from the trace of Einstein's equation:
\beq
 -{3\over 2} R =  g^{AB}G_{AB} =
g^{AB}\left[\de_A\Phi_0 \de_B\Phi_0 -{1\over2}g_{AB}\left(g^{CD}\de_C\Phi_0 \de_D\Phi_0 +
  V\right)\right]=a^{-2}\left(-{3\over 2}\Phi_0'^2-{5\over 2}a^2 V\right).
\eeq
}
 is given by:
\beq\label{b6}
R = a^{-2} \Phi_0'^2 + {5\over 3}V = -8 a^{-2} \left({a''\over a} +
 {1\over2}{a'^2 \over a^2}\right) = - {16\over 3}e^{4/3 B}\left(B'^2 - B''\right).
\eeq

In the rest of this paper we will consider space-times which have an
asymptotically $AdS_5$ region near $y=0$ ,
 where the metric and scalar field behave as
\beq
ds^2 \sim {1\over (k y)^2}\left(dy^2 + \eta_{\mu\nu}dx^\mu dx^\nu\right),
\quad \Phi_0' \sim 0,
\eeq
where $k$ is the $AdS$ curvature scale and it is determined by the value
of the scalar potential near $y=0$.
In the dual field theory  language this means that we are considering
 gauge theories that are UV-complete and do not necessitate a UV cut-off.

We will take the $y$ coordinate to range from zero to either infinity or
to a finite value $y_0$.  Far from
the $AdS$ boundary, the background scalar field will have a nontrivial
 profile, and the metric will generically differ from the one of $AdS$.

We can briefly compare this situation with what happens in the RS models,
 where the 5D metric  is $AdS$ but the space-time
 is cut-off by a $Z_2$ orbifold action at $y=y_c>0$. These models supports
 a normalizable massless spin-2 mode which is
 localized near $y_c$ and mediates 4D gravity. In our case instead,
 we take the space-time to extend
 all the way to the AdS boundary. As a consequence, the massless
RS mode becomes non-normalizable and is
 projected out of the spectrum. In the rest of the paper we will
investigate under what conditions the space-times under
  consideration still support a normalizable massless spin-2 mode,
which this time will be localized far from the $AdS_5$ boundary.

Recall that,  as explained e.g. in \cite{apr}, the RS models have a dual
 holographic interpretation as  a
4D CFT with a UV cut-off, living on the UV brane. According to this
 interpretation, 5D normalizable modes
that are localized near the UV brane correspond to elementary states
in the 4D theory, while modes that are
 localized far from the UV brane are non-elementary. In particular,
in the holographic interpretation of the
 RS models, the graviton is always treated as elementary,
and the Einstein action is part of the fundamental
 action defined at the UV cut-off.
In the holographic description of our setup  the graviton
is not part of the fundamental degrees of freedom, which consist
in an asymptotically conformal  field theory at strong coupling.
A massless spin-2 mode localized far from the boundary of $AdS$
is instead to be  interpreted as arising from the IR dynamics.

\section{Linear Fluctuations}

In this section we derive the equations for the  linear fluctuations
 around the background introduced above.
 We derive the  perturbation equations by varying  the action, eq. (\ref{action}),
 expanded up to second order in the fluctuation.
 This is useful to determine the normalization of the fluctuations.
 This kind of analysis has appeared in various works,
see e.g.  \cite{CEHS,Tanaka:2000er,Csaki:2000zn,peloso}. However, we
  found that some subtleties that arise when dealing with 4D-massless
modes have been overlooked, and this case necessitates separate treatment.
A detailed analysis of this case is performed in Appendix A.2.
The details of this procedure can be found in Appendix A,
here we will give only the salient points.

A generic perturbation of the metric and the dilaton
around the background (\ref{b0}) can be written as
\bea\label{fluc0}
 ds^2 && = a^2(y)\left(\eta_{AB} + h_{AB}\right) \nn\\
&& =a^2(y)\left[\left(1+2\phi\right)dy^2 + 2A_{\mu}dydx^{\mu}+
\left(\eta_{\mu\nu} + h_{\mu\nu}\right)dx^\mu dx^\nu\right], \\
\Phi && = \Phi_0(y) + \chi,
\eea
where $\phi$, $A_{\mu}$, $h_{\mu\nu}$ and $\chi$ are functions of $y,x^{\mu}$.


Under a 5D diffeomorphism, $(\delta y = \xi^5, \delta x^\mu = \xi^\mu)$,
the fluctuations defined by eq. (\ref{fluc0}) transform as:
\bea
&&\delta h_{\mu\nu} = -\de_\mu \xi_\nu - \de_\nu \xi_\mu
-2 \eta_{\mu\nu}{a'\over a} \xi^5, \label{gauge1}\\
&&\delta A_\mu = -\xi_\mu' -\de_\mu\xi^5,\label{gauge2}\\
&&\delta \phi = -\xi^{5'}-{a'\over a} \xi^5,\label{gauge3}\\
&& \delta \chi = -\Phi_0'\xi^5. \label{gauge4}
\eea

Due to gauge invariance, not all of these perturbations
 are dynamical: the usual counting of degrees of freedom for
gravitational theories implies that the metric and
dilaton fluctuation, ($h_{AB}$, $\chi$), contain 16 components, out
 of which 5 are eliminated by gauge transformation
and another 5 can be eliminated through the non-dynamical components
 of Einstein's equations. We are left with a total of
 $6$ degrees of freedom, which correspond to a 5D massless spin-2 plus a scalar.

Expanding eq. (\ref{action}) to second order  around the
background (\ref{b0}) we obtain the following action for the perturbations:
\bea\label{action2}
S^{(2)} &=&  {1\over 2k_5^2}\int d^4x dy\, a^3(y)
\Bigg[ L_{ein}^{(2)}  - {1\over4} h'_{\rho\sigma}h'^{\rho\sigma} + {1\over4} (h')^2
-{1\over4}F_{\mu\nu}F^{\mu\nu} \nn\\
& -&  \de_{\mu}\chi \de^{\mu}\chi -\chi'^2- {1\over2}a^2
\de^2_{\Phi}V \, \chi^2 -\de^{\mu}\phi\left(\de^{\nu}h_{\mu\nu} - \de_\mu h\right) \nn\\
&+& 2 \Phi_0' \phi' \chi + \Phi_0' h' \chi
+ 4 \Phi_0' \phi \chi' + 2 \Phi_0' A^{\mu}\de_{\mu}\chi \Bigg] \nn\\
&-& \left(a^3 A^{\mu}\right)' \left[\de_{\mu}h - \de^{\nu}h_{\mu\nu}\right] + \left(a^3 \right)' \left[ -2 A_{\mu}\de^\mu\phi - 2 \phi\phi' - \phi h'\right],
\eea
where from now on $h\equiv h_\mu^\mu$,
$F_{\mu\nu} = \de_\mu A_\nu - \de_\nu A_\mu$,  and
\beq\label{einstein}
 L_{ein}^{(2)} = -{1\over 4}\de^\mu h_{\rho\sigma}\de_\mu h^{\rho\sigma}
 + {1\over2} \de^\mu h_{\rho\mu}\de_\nu h^{\rho\nu}-
{1\over2} \de^\mu h \de_\rho h^{\rho\mu} +{1\over 4} \de^\mu h \de_\mu h,
\eeq
is the  quadratic part of the 4D Einstein-Hilbert Lagrangian.

\subsection{Classification of Physical Fluctuations}
Varying the action (\ref{action2}) w.r.t. $h_{\mu\nu}$,
$A_{\mu}$, $\phi$ and $\chi$, we obtain field  equations
 which correspond to the linearized Einstein-Dilaton
field equations, and are reported in  Appendix A.
We are interested in solutions that  correspond to
particles of a definite mass in four dimensions and
that are normalizable in the fifth dimension (finite 4D kinetic term after
dimensional ``reduction''), i.e. solutions of the form
\beq\label{separation}
\Omega(y,x^\mu) = \Psi_{k}(y) \omega_{k}(x^\mu),
\qquad \Box \omega_{k} = - k^2 \omega_{k},
\eeq
where $\Omega(y,x^\mu)$ is any of the fields under consideration.
Due to the linear character of the equations we
can consider each mode, labeled by 4-momentum $k_\mu$, separately.
 One has to separate two cases: massless modes
($k_\mu k^\mu =0$) and massive ones ($k_\mu k^\mu =-m^2 \neq 0$).
 This distinction  is needed if we want to classify
the modes according to transformation properties w.r.t. the 4D
  Lorentz group. This analysis is carried out in detail in Appendix A.

\vspace{0.3cm}
{\bf Massive Sector}
\vspace{0.3cm}

In the massive sector, the gauge-invariant propagating
fields are  a transverse, traceless  tensor $h_{\mu\nu}^{TT}$
(a massive spin-2 in 4D) and a scalar field $\zeta$,
for a total of ($5+1=6$) degrees of freedom.
The scalar field $\zeta$ is defined as
\beq\label{zeta0}
\zeta = \psi - {\chi \over z}, \quad \psi\equiv
 {1\over 6}\left(h_\mu^\mu - {\de^\mu\de^\nu h_{\mu\nu} \over m^2}\right),
\eeq
where $z(y)$ is defined in eq. (\ref{zed}).
The field equations are:
\bea
&& h^{TT''}_{\mu\nu} + 3 {a'\over a} h^{TT'}_{\mu\nu} + \Box_4  h^{TT}_{\mu\nu} =0, \label{tensormass}\\
&& \zeta'' + \left(3 {a'\over a} + 2 {z'\over z}\right) \zeta' + \Box_4 \zeta =0. \label{scalarmass}\eea
Notice that the tensor equation is the same as the massless scalar field eq. in the 5D  background (\ref{b0}).

\vspace{0.3cm}
{\bf Massless Sector}
\vspace{0.3cm}

In the massless sector, the six physical degrees of freedom are divided  as follows:  a transverse tensor and
vector, $  h_{\mu\nu}^{sTT}$ and $A^{sT}_{\mu}$, which are also \emph{spatially}-transverse (2 components each),
 plus two massless scalar fields $\zeta_1$ and $\zeta_2$.
In an appropriate gauge the corresponding fluctuations appearing in (\ref{fluc0}) take the form:
\bea\label{flucmassless}
&&h_{\mu\nu} =  h_{\mu\nu}^{sTT} + 2\eta_{\mu\nu}\psi, \quad \\
&&  A_\mu =  A_\mu^{sT} ,
\eea
with the transverse vector and tensor obeying:
\bea
&& h_{\mu}^{sTT\mu} =  \de^\mu h_{\mu\nu}^{sTT}=  \de^i h_{i\nu}^{sTT}= 0,\nn\\
&&  \de^\mu  A_\mu^{sT} = \de^i  A_i^{sT} =0.
\eea
The field equations for these fields and the remaining scalar fluctuations $\phi$ and $\chi$ are:
\bea
&& (h_{\mu\nu}^{sTT})'' + 3 {a'\over a} (h_{\mu\nu}^{sTT})' = 0; \label{tensor00}\\
&& (a^3 A_\mu^{sT})' = 0; \label{vectorT}\\
&& \phi = -2\psi; \\
&& \zeta_1' = 0, \qquad  \left({a^4\over a'} \zeta_2 \right)' = -2 a^3 \zeta_1,\label{scalar00}
\eea
where $\zeta_1$ and $\zeta_2$ are given by the linear combinations:
\beq\label{zeta12}
\psi = \zeta_1 + \zeta_2, \qquad \chi = z \zeta_2,
\eeq
and we have of course  $\Box_4  h_{\mu\nu}^{sTT} = \Box_4 A_\mu^{sT} = \Box_4 \zeta_1 = \Box_4 \zeta_2 =0$.


\subsection{Effective Actions and Normalizability}

Next, we determine the normalization conditions for the modes found in the previous section. To do this we
 substitute in the action (\ref{action2}) the fields whose $y$-profile satisfies the field equations for a given mass.

\subsubsection{Tensor Modes}

The equation obeyed by  the transverse traceless tensors modes (both in the massless and massive case) is  eq (\ref{tensormass}):
\beq
h^{TT''}_{\mu\nu} + 3 {a'\over a} h^{TT'}_{\mu\nu} + \Box_4 h_{\mu\nu}^{TT} =0.
\eeq
Separating $h^{TT}_{\mu\nu}$ into a purely spatial part and a $y$-dependent profile:
\beq
 h_{\mu\nu}=h(y)h^{(4)TT}_{\mu\nu}(x),\qquad  \Box_4 h^{(4)TT}_{\mu\nu}= m^2 h^{(4)TT}_{\mu\nu},
\eeq
we get the equation for the profile $h(y)$:
\beq\label{tensor2}
h'' + 3 {a'\over a} h' + m^2 h =0.
\eeq
In the massive case there are five independent polarizations, while in the massless case we gauged away
all but two polarizations (the ones of a physical massless spin-2 in 4D).

It is convenient to introduce the function $B(y)$ and the wave-function $\psi_t(y)$:
\beq\label{B}
a(y)= e^{-{2\over3} B(y)}, \qquad \psi_t(y)=  e^{-B(y)}h(y).
\eeq
Then (\ref{tensor2}) becomes a Schr\"odinger-like equation for $\psi_t(y)$:
\beq\label{sc2}
-\psi_t'' + V_t(y) \psi_t = m^2 \psi_t, \qquad V_t = B'^2 -B''.
\eeq
Notice that the potential $V_t$ is proportional to the 5D  curvature scalar, see eq. (\ref{b6}):
\beq
V_t = -{3\over 16}a^2 R.
\eeq

To get the normalization measure, we insert the ansatz  $h_{\mu\nu}=h(y)h^{(4)TT}_{\mu\nu}(x)$ into (\ref{action2}):
\bea
S^{TT} &=& {1\over 2k_5^2}\int d^4x dy\, a^3(y) \left[ -{1\over4} h(y)^2 \de_\rho  h^{(4)}_{\mu\nu} \de^\rho
 h^{(4)\mu\nu} -{1\over 4} h'(y)^2  h^{(4)}_{\mu\nu}   h^{(4)\mu\nu}\right]\nn\\
&=&  {1\over 2k_5^2}\int dy\, a^3(y) h(y)^2 \int  d^4x \, \left[ -{1\over4} \de_\rho  h^{(4)}_{\mu\nu} \de^\rho
 h^{(4)\mu\nu} - m^2   h^{(4)}_{\mu\nu}   h^{(4)\mu\nu}\right].
\eea
In the second step we have integrated by parts and used (\ref{tensor2}).
From the expression above it is clear that the mode corresponds to a normalizable 4D mode if $a^{3/2} h(y)$ is
 square-integrable or, in terms of the wave function defined by (\ref{B}), if
\beq\label{tensornorm}
 \int  dy\,|\psi_t(y)|^2 < \infty.
\eeq
That is, the solution of eq. (\ref{sc2}) must be normalizable in the usual quantum-mechanical norm.

\subsection{Vector Modes}

In this sector  there are only zero-modes, and no massive ones. In the gauge we are using, the transverse
vector is purely contained in $A_\mu$, since we gauged away the vector mode from $h_{\mu\nu}$.
From eq. (\ref{vectorT}) we obtain the profile of $A_\mu$:
\beq\label{vectorsolution}
A_\mu(x,y) = {1\over a^3(y)} A_\mu^{(4)}(x).
\eeq
From the  action (\ref{action2})  we obtain the normalization condition:
\beq
S^{(V)} = {1\over 2k_5^2} \int d^4x dy\, a^3(y) \left[-{1\over 4} F_{\mu\nu}F^{\mu\nu}\right] = {1\over 2k_5^2}
 \int  dy\, a^{-3}(y)  \int d^4x   \left[-{1\over 4} F_{\mu\nu}^{(4)}F^{\mu\nu{(4)}}\right]
\eeq
So we have a normalizable 4D massless vector if
\beq\label{vector_norm}
\int  dy\, a^{-3}(y) < \infty
\eeq

\subsection{Scalar Modes - Massive Case}

The relevant scalar  quantity is $\zeta(y,x)$, defined in (\ref{zeta0}). In fact, when  we use
an ansatz that solves the 5D equations, $\zeta(y,x)$ is the only scalar mode that appears in the actions.
 Decomposing $\zeta(x,y)$ in modes with a given non-zero  4D mass we get, from eq. (\ref{scalarmass}), the equation for the profile:
\beq
 \zeta'' + \left(3 {a'\over a} + 2 {z'\over z}\right) \zeta' + m^2 \zeta =0.
\eeq
Defining:
\beq\label{G}
C(y) = \log z(y), \qquad \psi_s(y) = e^{-G(y)}\zeta(y), \qquad G = B - C,
\eeq
we get a Schr\"odinger  equation for the wave-function $\psi_s$, like eq. (\ref{sc2}), except for the substitution $B(y)\to G(y)$ in the potential:
\beq\label{sc0}
-\psi_s'' + V_s(y) \psi_s = m^2 \psi_s, \qquad V_s = G'^2 -G''.
\eeq

As was shown in \cite{peloso}, the quadratic action in the scalar sector reduces to an action
for the gauge-invariant field $\zeta$ alone:
\beq
S^{(S)} =  {1\over 2k_5^2}\int d^4x dy\, a^3(y) z^2(y) \left[\zeta'^2 + (\de_\mu\zeta)^2\right]
= {1\over 2k_5^2}\int d^4x dy\, e^{-2G(y)} \left[\zeta'^2 + (\de_\mu\zeta)^2\right].
\eeq
It follows that , as in the case of the tensor modes, the normalization condition in terms of $\psi_s(y)$ involves the trivial measure:
\beq\label{scalarnorm}
 \int  dy\,|\psi_s(y)|^2 < \infty.
\eeq

\subsection{Scalar Modes - Massless Case}

In the gauge used in Section 3.1 and discussed in detail in  Appendix A.2,
where the tensor and vector massless fluctuations propagate
two physical components each,
the scalar metric fluctuations appearing in eq (\ref{fluc0}) are given by:
\beq\label{fluc3}
\chi, \quad h_{\mu\nu}= 2\eta_{\mu\nu}\psi, \quad \phi = -2\psi, \quad A_\mu=0.
\eeq
The linear combinations
$\zeta_1= \psi- \chi/z$ and $\zeta_2=z \chi $ satisfy  eqs. (\ref{scalar00}).
  These equations can be easily solved in terms of two independent integration ``constants'' $F(x), G(x)$,
 which play the role of the two 4D scalar massless degrees of freedom :
\beq\label{scalarsolution}
\zeta_1 = F(x); \qquad \zeta_2 = {a' \over a^4}(y) G(x) -2 \left({a'\over a^4}\int  a^3 \right)(y) F(x).
\eeq

 To see how these modes are normalized in 4D terms, we insert  (\ref{fluc3}) in the action (\ref{action2}),
 and after  using the appropriate zero-mode scalar field equations, found
in Appendix A.2 (namely  eqs.
 (\ref{yy0}), (\ref{dilaton0}), (\ref{scalarzero}) and (\ref{scalar1})) ,
  the action reduces to a purely 4D kinetic term:
\beq\label{kinetic1}
S^{(S)}_{kin} = {1\over 2k_5^2}\int d^4x dy\, a^3(y)\left[  -6 \de^\mu \psi \de_\mu \psi     - \de_{\mu}\chi \de^{\mu}\chi \right].
\eeq
The fields appearing in eq. (\ref{kinetic1}) are not yet canonically normalized in four dimensional terms, since they
have a nontrivial $y$-dependence. To determine the actual normalization we must use the explicit form of the solution,
 eqs. (\ref{scalarsolution}). The result in terms of the $y$-independent fields $F(x), G(x)$ is not particularly
  illuminating. In the next Section
we will use only its asymptotic forms close to the
boundaries of space-time to determine whether the
   corresponding modes are indeed normalizable.

\section{Hunting for Zero-modes}

After setting the stage with the analysis of the previous section,
we proceed to the main purpose of this paper: asking whether it is possible,
in this setup, to have  modes that resemble massless 4D gravitons,
i.e. normalizable solutions of eq. (\ref{tensor2}) with $m^2=0$.
If the answer is positive, at energies below the mass of the first
massive excitation we can truncate the theory to an effective 4D theory
of the massless states, by integrating over $y$.
The requirement of normalizability is equivalent to asking that these modes have a
finite 4D kinetic term (a finite value of the four-dimensional Planck scale)
 after the $y$-coordinate has been integrated out.
Moreover, we would like to have  zero-modes of spin-2 only, with no additional
 massless vectors or scalars.

The spectrum of fluctuations depends essentially on the form of the scale factor,
described by  $B(y)$. This function determines the
potential entering the tensor and scalar perturbation equations, eqs. (\ref{sc2})
and (\ref{sc0}), as well as the respective normalizability conditions.
The scale factor   depends on the solution of the background Einstein's equations,
that in turn depends on the dilaton  potential,
 entering the original action, (\ref{action}).  One could ask which kind of
 dilaton potentials generate solutions with a spectrum like
 the one we are looking for.  However, in the 5D Einstein-Dilaton theory
there is a huge simplification \cite{freedman}: under certain
 conditions, discussed in the next subsection, given an arbitrary function
 $a(y)$, we are automatically  guaranteed that there exists a
 dilaton potential $V_a(\Phi)$ such that a space-time with scale factor
 $a(y)$ is a solution of Einstein's equations. Therefore we can parametrize
  our model by a choice of $a(y)$, or equivalently $B(y)$,
and reverse-engineer $V(\Phi)$ at the end.  This is much simpler, since the potentials
  in eqs.  (\ref{sc2}) and (\ref{sc0}) and the normalization
conditions are express directly as functions of the scale factor. At the end, we have
  to check that the solution we obtained does not contain pathologies,
e.g. violate some  positive energy condition.

To summarize, the strategy we follow is the following: we investigate
in what geometries,  parametrized by  $B(y)$, the spin-2 Schr\"odinger
equation, (\ref{sc2}), admits normalizable zero-energy eigenstates,
and at the same time the spin-0 and spin-1 massless modes, given  by eqs.
(\ref{scalarsolution}) and (\ref{vectorsolution}), respectively, are
\emph{not} normalizable.

We will restrict our analysis to asymptotically $AdS_5$ space-times,
since in this case our setup can be given a holographic interpretation
in terms of an asymptotically conformal 4D field theory. If we relax
this assumption, an example of a spectrum like the one we are looking for
can be found e.g. in the RSII model\cite{RS}: there is no dilaton,
just a cosmological constant, and the  scale factor and tensor potential take the form
\beq
B(y) = {3\over 2}\log \left( \e + k|y|\right), \quad V_t(y) = {15\over4}
{k^2\over (\e + k|y|)^2} - {3\over 2}k \d(y)
\eeq
This potential supports a massless zero-mode graviton with
wave-function peaked around $y=0$, given by (as we will see in the next subsections):
\beq\label{wfRS}
\psi_{RS} (y) = e^{-B(y)} = {1\over (\e + k|y|)^{3/2}}
\eeq
 This corresponds, in the holographic interpretation, to an
 elementary graviton, added to the  CFT (cut-off at an UV energy scale
 $k/\e$) in the UV of the theory. In our case, we  are more
 interested in the possibility of realizing the graviton as a non-elementary
 state of the CFT, therefore we need the wave-function
to vanish in  the UV. In fact, if we  remove the UV cut-off
from the RS model, $\e \to 0$, and let the space-time extend all
the way to the $AdS_5$ boundary, we can see immediately from eq. (\ref{wfRS}) that the  RS zero-mode becomes non-normalizable in the
 norm  (\ref{tensornorm}). On the other hand, the second
independent zero-energy  solution of eq. (\ref{sc2}) behaves near the boundary
  as $y^{5/2}$, and can  give rise to a normalizable zero mode,
depending on the
large $y$ behavior of the scale factor. We will analyze this
question in detail in the rest of the paper. We classify the geometries
 according to their properties at large $y$:  the range of $y$
is either infinite,  or the $y$ coordinates stops at some ``end of space''
 point $y_0$, where there could be either a singularity or a
boundary. Before starting this analysis of the zero-modes, we briefly review
 what is the degree of arbitrariness we can use in choosing
the geometry through the function $B(y)$.

\subsection{Energy Conditions and  Constraints on $B(y)$}\label{constraint}

One of the features of the 5D model we are considering is that
one can give a general classification of the possible
behavior of fluctuations for the metrics of the form (\ref{b0})
without having first to specify the dilaton potential:
the spectrum  depends on a single function $B(y)$, that
can be chosen fairly arbitrarily. In fact,
given any choice of scale factor $a(y)$, or equivalently any
function $B(y)$, so long as the condition:
\beq\label{positive}
\Phi_0'^2 = 2B''+ {4\over 3}B'^2 > 0,
\eeq
is satisfied, we can  always find a potential $V(\Phi)$ such
that Einstein's equations  are solved by that choice
of $B(y)$\cite{freedman}. To see this, do a coordinate
transformation so that the metric takes the form:
\beq
ds^2 = dr^2 + e^{-4B(r)/3}\eta_{\mu\nu}dx^\mu dx^\nu,
\qquad {d y\over d r} = e^{2B(y)/3}.
\eeq
Then (\ref{b4}),(\ref{b5}) can be rewritten as (a dot denotes derivative w.r.t. $r$):
\beq\label{b7}
\dot{\Phi}_0^{2}(r) = 2\ddot{B}(r), \qquad V(\Phi_0(r))=
2\ddot{B}(r) - {16\over3}\dot{B}^2(r).
\eeq
If $\dot{\Phi}_0(r)\neq 0$, $\Phi_0=\Phi_0(r)$ can be
inverted to give $r=r(\Phi_0)$, and  we can define a ``superpotential''
$W(\Phi)=(4/3)\dot{B}(r(\Phi))$. From (\ref{b7}) we then
obtain $V$ as a function of $\Phi$ as
\beq
V(\Phi)={9\over 4} \left(\de W\over \de\Phi\right)^2 - 3W^2.
\eeq
If $\dot{\Phi}_0$ vanishes at some isolated points, we can
repeat the above procedure piecewise in every range of
 the $r$ coordinate in which   $\dot{\Phi}_0(r)\neq 0$. So,
if needed we can relax the condition (\ref{positive})
to include the equality sign.
Thus, in what follows we can think of $B(y)$ as a function
that we can choose at will, provided  (\ref{positive})
is satisfied in the whole range of $y$ under consideration.

From the first of eqs. (\ref{b7}), we see that the condition
 (\ref{positive}) is equivalent to asking that the
 function $B(r)$  have non-negative second derivative.
In terms of $B(y)$ this means that
\beq\label{positive2}
 \textrm{\emph{the function}} \quad \dot{B}(r) \quad \textrm{\emph{must be non-decreasing}}.
\eeq
In terms of the conformal coordinate $y$ that we use throughout this
paper, the function $\dot{B}(r)$ is given explicitly by:
\beq\label{positive3}
\dot{B}(r) = B'(y)\dot{y} =B'(y)\exp[2B(y)/3].
\eeq

Notice that condition (\ref{positive}) is nothing but the Null
Energy Condition (NEC) for the space-times in consideration. The
dilaton stress tensor is
\beq\label{st}
T_{AB} = \de_A \Phi \de_B\Phi - {1\over 2} g_{AB}
\left[\de^C\Phi\de_C\Phi + V(\Phi)\right].
\eeq
The NEC requires that for any null vector $v^A$ ,
$T_{AB} v^A v^B \geq 0$ (see e.g. \cite{HE}). From eq. (\ref{st}) this means,
for our background $\Phi=\Phi_0(y)$,
\beq
(v^y)^2(\Phi')^2   \geq 0.
\eeq
In order to obtain more general space-times that violate
the NEC (if one were willing to do so), one would have to include
``ghost'' scalar fields, with opposite sign kinetic term.
 In this case the sign of the first term in (\ref{st})
would be negative,  and one would be able to consider
space-times in which (\ref{positive}) is not satisfied.
We will not consider this possibility any further in this work.

Let us look at the NEC in the case of asymptotically $AdS$
space-time. Here,  $ds^2 \sim y^{-2}(dy^2 +dx_\mu^2)$ as $y\sim 0$,
and we have
\beq
B(y) \sim {3\over 2}\log y, \quad \dot{B}(0) \sim {3\over2}.
\eeq
In particular, $B'(y)$ cannot  vanish at any finite positive
value of $y$, otherwise $\dot{B}$ would also
 vanish there (see eq. (\ref{positive3})) and it
would  decrease from $3/2$ to zero. Since $B'(0)>0$,
 this implies that  $B'(y)>0$ for any finite $y$, i.e. $B(y)$
must be monotonically  increasing. This is a necessary
 condition for the NEC to be satisfied.

One may wonder whether one could construct more general
5D background, using
different kinds of bulk matter than just a single scalar field.
For example, one could consider a generic
5D ``perfect fluid'' , with stress tensor of the form
(if we limit ourselves to space-times that preserve 4D Lorentz
invariance)
\beq\label{fluid}
T_{AB}= \left(\begin{array}{cc}W & \\
                                 &  T\eta_{\mu\nu}
\end{array}\right).
\eeq
Different ``equations of state'' T = T(W) give rise to different
solutions, exactly as in cosmological models.
Einstein's equations are
\beq
6\left(a'\over a\right)^2 = W, \qquad 3 {a''\over a} = T,
\eeq
 and the conservation law  obtained from these two equations is
\beq
W' +  2 {a'\over a}\left(W - 2T\right) =0.
\eeq
For example, a linear relation  $T = \omega W$ gives rise
to power-law solutions, $W \propto a^{2(2\omega -1)}$ and
$a(y) \propto y^{1/(1-2\omega)}$. Patching together different
solutions in different $y$ regions, we can obtain
 piece-wise continuous bulk geometries, with a different
power-law in each region. We can also do the same thing smoothly,
by adding various different components characterized by
different $\omega$'s: we would have transitions between different
regions in each of which a different component dominates.
From eq. (\ref{fluid}) one can see that the null energy condition
requires $\omega <1$.  Matter
that violate this inequality is analogous to ``phantom-like''
matter in usual 4D cosmology, one particular realization being
the ``ghost'' dilaton we mentioned above.

The parametrization (\ref{fluid}) is in principle more general
than having  just a dilaton\footnote{Dilaton matter is described by
a stress tensor of the form (\ref{fluid}) with:
\beq
W= {1\over 2} \left((\Phi')^2 - a^2 V(\Phi)\right), \quad T =
 - {1\over 2} \left((\Phi')^2 + a^2 V(\Phi)\right). \nn
\eeq
}. However, the equations  for the tensor fluctuations,
eq (\ref{tensormass}),  depend \emph{only on the scale factor},
and we have seen that already with a single scalar field
 we can obtain \emph{any} warped metric of the form (\ref{b0})
(therefore the most general $SO(3,1)$-preserving 5D metric)
that does not violate the NEC. Therefore
the results we obtain in the Einstein-Dilaton model for the
spectrum of spin-2 modes  are completely model-independent:
they depend only on the geometry of the solution, no matter
what is the specific matter or field content of the model.
These details  have an influence on the scalar sector of the
fluctuations, whose corresponding equations (\ref{scalarmass}),
(\ref{scalar00}) contain the non-universal function $z(y)$.

\subsection{Spin 2}

We first look for spin-2 zero-modes, i.e. a transverse traceless tensor
with two propagating d.o.f.,
whose profile satisfies eq. (\ref{sc2})  with $m^2=0$:
\beq\label{sc2-zero}
-\psi_t'' +  \left( B'^2 -B''\right)\psi_t = 0.
\eeq
The two independent solutions are:
\beq
\psi_t^{UV}(y) = e^{-B(y)}, \qquad   \psi_t^{IR}(y) =
e^{-B(y)}\int dy'\, e^{2B(y')},\label{zero2}
\eeq
where the labels indicate the region where the wave-function is peaked.
The question of finding zero modes reduces to asking whether any  of these solutions is normalizable
 for a given choice of the function $B(y)$.

In an asymptotically $AdS$ space-time, $B(y) \sim {3\over2} \log ky$ as $ y\sim 0$, and
the two tensor modes  behave for small $y $ as:
\beq
\psi_t^{UV}(y) \sim y^{-3/2}, \qquad \psi_t^{IR}(y) \sim y^{5/2}.
\eeq

This immediately makes  $\psi_t^{UV}$ non-normalizable, and excludes it from the spectrum.
Notice that this mode is the one that survives in RSII, where $AdS$ is cut-off before reaching the boundary.
 In the holographic description this is the mode that is relevant in the $UV$,
and couples as a source to the CFT stress tensor. The other mode instead vanishes close to the boundary,
 and can be interpreted holographically as a mode that arises in the IR.
 The constant of integration in  $\psi_t^{IR}(y)$ is fixed by normalizability around $y=0$,
so that the correct  definition of the profile of this mode is:
\beq\label{zero2-2}
\psi_t^{IR}(y) = e^{-B(y)}\int_0^y dy'\, e^{2B(y')}.
\eeq

Whether this function  is normalizable or not depends uniquely  on the behavior  of $B(y)$ for large $y$.
There are two possibilities:
\begin{itemize}
\item the $y$-coordinate extends to $+\infty$.
\item space-time ends  at a finite value  $y=y_0$.
\end{itemize}
In the first case the resulting space-time is  regular, in the second case
there is either a singularity or a boundary at $y=y_0$.

\subsubsection{Regular Space-times}

Suppose the $y$-coordinate ranges from $0$ to infinity. Then one can show immediately  that the wave-function (\ref{zero2-2})
is not normalizable,  whatever is the large $y$ behavior of $B(y)$: $\psi_t^{IR}(y)$ always diverges as $y\to +\infty$:
\begin{itemize}
\item if $B(y) \to const.$\footnote{this case includes  $B(y) \to 0$},   $\psi_t^{IR}(y) \sim y$
\item if $B(y) \to -\infty$, then $\int_0^{y} \exp[2B(y')]$  must either diverge to $+\infty$ or go to a
 {\em strictly positive} constant as we take $y\to +\infty$, since it is a monotonically increasing function
 starting at zero at $y=0$. In either case, the product of that integral with $\exp[-B(y)]$ diverges at $+\infty$.
\item If  $B(y) \to +\infty$ then the integral of $\exp[2B(y')]$ diverges much faster as $y\to+\infty$ than the
pre-factor  $\exp[-B(y)]$ goes to zero, making the product diverge.
\end{itemize}
These arguments show that the only case in which one can have a massless spin-2 in 4D is if the space
 terminates at some $y=y_0$. This is the case  we will consider in the rest of this section.


\subsubsection{Singular Spacetimes}

The ``end of space'' singularity can be caused by $B(y)$ going to $+\infty$
(space-time shrinks to zero size)  or to $-\infty$ (the scale factor blows up)
at $y=y_0$, or by a divergence in $B'$ or $B''$ while $B(y_0)$ stays finite
  (in this case the metric  remains finite but the curvature blows up).
 We will analyze these cases separately.

\begin{itemize}
\item {\bf $B(y)$ diverges as power-law}

Suppose that around some $y_0>0$
\beq
B(y) \sim {c \over(y_0-y)^{\b}}, \quad \b>0
\eeq
where $c$ is a constant that can be positive or negative. Let us first check what constraints we get from condition
 (\ref{positive}): around $y_0$, we have:
\beq
\dot{B}\sim {c \b \over(y_0-y)^{\b+1}}e^{(2c/3) (y_0-y)^{-\b}},
\eeq
where we have used the expression of $\dot{B}$ appearing in eq. (\ref{positive3}).
Suppose  $c<0$: then  $\dot{B} \to 0$ as $y\to y_0$. Since around $y=0$ $\dot{B} \sim 3k/2 >0$, $\dot{B}$
cannot be  increasing everywhere, thus violating the  condition (\ref{positive2}). So we must take $c>0$.
But in this case one can immediately see that  $\psi_t^{IR}(y)$ is not normalizable around $y_0$, since again the integral in (\ref{zero2-2}) diverges faster than the prefactor goes to zero.

\item {\bf $B(y)$ diverges logarithmically}

Consider now the case that $B(y)$ diverges logarithmically:
\beq\label{logwarp}
B(y) \sim -\a \log(y_0-y), \quad y\to y_0.
\eeq
Now we have
\beq
\dot{B} \sim \a (y_0-y)^{-2\a/3-1},  \quad y\to y_0.
\eeq
We must require $\a>0$ in order to ensure that $\dot{B}$ does not decrease in some region. In this case
 we do find normalizable zero-modes in a certain range of $\a$: around $y_0$ we have:
\beq
\psi_t^{IR}(y) \sim (y_0-y)^\a \left( const + (y_0-y)^{-2\a+1}\right)
\eeq
which is normalizable if $\a<3/2$.
In this case, however, the spectrum of the model is not completely specified,
and the existence of the zero mode requires special boundary conditions to be imposed at the singularity.
 This will be discussed in the next section.

With the warp factor behaving as in  (\ref{logwarp}), the graviton potential $V_t$ at $y_0$ behaves as
\beq
V_t = B'^2 - B'' \sim {\a^2-\a\over (y_0 -y)^2} \qquad y\sim y_0
\eeq
so it goes to $+\infty$ if $1<\a<3/2$, and to $-\infty$ if $0<\a<1$. According to eq. (\ref{b6}),
the 5D scalar curvature goes like
\beq\label{curvature}
R \propto - e^{4/3\,B}V_t \sim - \,{\a^2-\a\over (y_0 -y)^{(2+4\a/3)}}
\eeq
 so it goes either to $-\infty$ ($1<\a<3/2$) or $+\infty$ ($0<\a<1$). For the limiting case $\a=1$ one
  must look at the subleading behavior at the singularity (the potential will diverge less dramatically,
  as $(y_0-y)^{-1}$).

\item {\bf $B(y)$ finite}

There is another possibility, namely that the space terminates at a point $y=y_0$ where the scale factor
is finite but one of its derivatives diverges. This is still a curvature singularity, as one can see from eq.
(\ref{b6}), and it means that we have to terminate our space-time at this point.

Suppose that  as $y\sim y_0$
\beq\label{finite}
B(y) \sim B_0 + c (y_0-y)^{\b} \qquad \b>0
\eeq
In order to have a singularity, there must be some derivative of $B$ that blows up at $y_0$,
so we will assume $\b<2$.

In this case the tensor zero-mode (\ref{zero2-2}) is clearly normalizable, as it approaches a finite value as $y\to y_0$.

\end{itemize}

\subsubsection{Regular Spacetimes with IR Boundary}

A final possibility that we have to mention is the  case  where there is no singularity at all,
but rather there is a boundary at an arbitrary position $y=y_0$ where the space is cut-off. In this case
all the modes of various spins that grow in the IR are automatically normalizable, and the spectrum is fully
determined by the boundary conditions one imposes at $y=y_0$. This case avoids all the problems related to having to
resolve the singularity, which is something that one eventually will want to do. However, from the point of view of the
 dual theory, the presence  of a boundary at an arbitrary position means that we put an IR cut-off at an arbitrary energy
 scale. This IR  modification   is something that cannot be reconstructed from the knowledge of the fundamental theory and
 its dynamics, which on the other hand are encoded in the bulk geometry. On the other hand, the space ending at a
 singularity also signals some strong IR dynamics in the dual theory, but this  IR modification is purely a consequence
 of the ``microscopic'' dynamics: the presence and the location of the singularity can be deduced from the defining
  parameters of the theory, which on the gravity side are encoded in the bulk dilaton potential and UV boundary
  boundary conditions for the dilaton and the scale factor.  This is the same reason why the authors of
  \cite{csaki-reece} use singular spaces rather than cut-off branes in the holographic approach to QCD:
  the singularity is seen as a dynamical, rather than \emph{ad hoc}, way to terminate the space and having
some nontrivial IR dynamics.




\subsection{Spin 1}

In this sector, according to eq. (\ref{vectorT}) we have a  vector of  the form:
\beq
A^T_\mu(x, y) = A^{(4)}_\mu (x) \,a^{-3}(y).
\eeq
The actual presence of this mode as part of the spectrum depends on whether or not the normalization
 condition (\ref{vector_norm}) is satisfied.


\subsubsection*{UV behavior}

In the case we are considering, where the space has an asymptotic $AdS_5$-like boundary at $y\to 0$,
we have $a^{-3}(y) \sim y^3$ as $y\sim 0$, so the integral in eq.  (\ref{vector_norm}) converges in the
UV. To see whether the vector mode is there or not we must consider what happens in the IR.

\subsubsection*{IR behavior}

In the cases where we found a normalizable massless spin-2 mode, i.e. the space has a curvature singularity at $y=y_0$,
let us see what happens to  the spin-1 normalization condition in the infrared:

\begin{itemize}
\item {\bf $B(y)$ diverges logarithmically}

We have seen in the discussion of the spin-2 IR behavior that positivity of the dilaton kinetic energy and
 normalizability of the spin-2 zero-mode  requires that close to $y_0$:
\beq
B(y) \sim -\a \log (y_0-y), \qquad 0<\a < 3/2.
\eeq
The spin-1 normalization condition,  (\ref{vector_norm}), becomes:
\beq
\int^{y_0} dy \, {1\over (y_0 -y)^{2\a}} \, < \infty,
\eeq
that is,   $\a<1/2$. So,  in the range $1/2<\a < 3/2$ we find a spin-2 zero-mode
but no spin-1 zero modes.

\item {\bf $B(y)$ finite}

In this case, condition  (\ref{vector_norm}) is clearly satisfied, and the massless spin-1 is normalizable.

\end{itemize}

\subsection{Spin 0}

There are two  independent scalar zero-modes, parametrized in terms of  the two
functions $F(x)$ and $G(x)$ introduced in Section 3.5. From eqs. (\ref{zeta12}) and (\ref{scalarsolution}), the independent metric and dilaton scalar fluctuations are given by:
\bea
&&\psi = \left(1 - 2 {a'\over a^4}(y) \int^y a^3\right)F(x) + {a'\over a^4}(y) G(x), \label{FG1}\\
&& \chi = z(y){a'\over a^4}(y) G(x) - 2z(y)\left( {a'\over a^4}(y) \int^y a^3\right)F(x).\label{FG2}
\eea
To check normalizability we have to insert the above expressions in eq. (\ref{kinetic1}).

\subsubsection*{UV behavior}

We first check which of the two modes $F,G$ is normalizable around $y=0$. For small $y$, $a(y) \sim y^{-1}$ and
 eqs. (\ref{FG1}), (\ref{FG2}) become:
\beq\label{FG3}
\psi(x,y) \sim -3 F(x) + y^2 G(x), \qquad \chi(x,y) \sim - 4z(y) F(x) + y^2 z(y)G(x)
\eeq
It is useful to write the expression for $z(y)$ directly in terms of $B(y)$. From eqs. (\ref{zed}) and (\ref{b4}) we have:
\beq\label{zed2}
 z^2(y) = {9\over 4}\left({2B''\over B'^2} + {4\over 3}\right).
 \eeq
In pure $AdS$ we have $3B'' = -2B^{'2}$, and $z(y)$ vanishes identically. Therefore,  the small $y$ behavior of the  depends on the subleading terms
in $B(y)$. Assuming, for small $y$,
\beq
B(y)\sim 3/2 \log k y + c y^{\gamma}, \quad  \gamma\geq 0
\eeq
 where  $c$ a constant, we find
\beq
z(y) \sim  y^{\gamma/2}.
\eeq

 Inserting eqs.  in the reduced action, (\ref{kinetic1}), we see that the dominant contribution to the integral at $y=0$
 has the form
\beq
S[F,G] \sim \int_0 d y {1\over y^3} \left[-54 (\de_\mu F)^2 + 36 y^2 \de_\mu F \de^\mu G - 6 y^4 (\de_\mu G)^2\right],
\eeq
from which we see immediately that the $F$-mode is non-normalizable in the UV, whereas
 the $G$-mode is.


\subsubsection*{IR Behavior}

Having eliminated  the $F$-mode due to its UV non-normalizability, we are left with fields $\chi$ and $\psi$ of the form
\beq
\psi  = {a'\over a^4}(y) G(x), \qquad \chi = z (y) {a'\over a^4}(y) G(x)
\eeq

In what follows  we consider only the cases where we found a normalizable spin-2 zero-mode to exist.

\begin{itemize}
\item {\bf $B(y) \sim -\a \log (y-y_0), \quad 0<\a<3/2$}

In this case, close to $y=y_0$,  $z(y)\sim const$ and we find:
\bea
S[G] &&\simeq \int^{y_0} d y a^3 \left(a'\over a^4\right)^2  (\de G)^2 \sim \int ^{y_0}(y_0-y)^{2\a}
\left[ {(y_0-y)^{4/3\a -2}\over (y_0-y)^{16/3\a}} (\de G)^2\right] \nn\\ &&   \sim \int^y_0 d y {(\de G)^2\over (y_0-y)^{2\a+2}}.
\eea
The integral diverges and the $G$-mode is not normalizable in the IR.
\end{itemize}

To treat the situation when the scale factor at $y_0$  is finite but its first or second derivative diverge,
 it is useful to consider two separate cases:

\begin{itemize}
\item {\bf $B(y) \sim B_0 + B_1 (y_0-y) + c (y_0-y)^\b, \quad  1<\b<2$}

In this case, close to $y_0$,  $a'/a \propto B_1$ is finite\footnote{We cannot have  $B_1 = 0$
since this would imply $B'\to 0$ as $y\to y_0$ which, together  with the finiteness of $B(y_0)$,
 violates  condition (\ref{positive2})}   but $a''/a$ diverges. From (\ref{zed}) and (\ref{b4})
 we  see that the function $z(y)$ can be written as:
\beq
z^2 = 6 - {a'' \,a\over a^{'2}} \sim (y_0-y)^{\b-2}, \qquad y\sim y_0,
\eeq
and  the effective action for the mode $G(z)$ is
\beq
S[G] \simeq \int^{y_0}d y a^3 \left(a'\over a^4\right)^2 z^2(z)  (\de G)^2  \sim \int^{y_0}d y (y_0-y)^{\b-2} (\de G)^2,
\eeq
which is finite at $y_0$ since we are assuming $\b<2$. Therefore in this case one of the two scalar modes is normalizable.

\item {\bf $B(y) \sim B_0 + c (y-y_0)^\b$, $0<\b<1$}

In this case both   $a'/a$ and  $a''/a$ diverge. Now  we have:
\beq
z(y)^2 \sim (y_0-y)^{-\b},
\eeq
and the effective action is again
\beq
S[G] \simeq \int^{y_0}d y a^3 \left(a'\over a^4\right)^2 z^2(z)  (\de G)^2  \sim \int^{y_0}d y (y_0-y)^{\b-2} (\de G)^2,
\eeq
but  the last integral diverges, since $\b<1$, and there are no normalizable scalar modes in this case.
\end{itemize}

\subsection{Summary and Discussion} \label{summary}

The results of this section are summarized in Table 1. As announced in the introduction, we found
that only if the fifth dimension terminates at a finite value of $y$ it is possible to have normalizable
graviton zero-modes.
In this case, the spectrum is purely discrete, since the tensor Schr\"odinger's equation (\ref{sc2})
is defined on the interval $(0,y_0)$. Therefore,  there is  a mass gap (generically of order $y_0^{-1}$)
that separates the zero-mode
from the massive KK modes.
This is unlike what happens e.g. in RSII \cite{RS}, where the KK tower is a continuum starting at $m^2=0$. To have a
phenomenologically acceptable scenario, the masses of KK gravitons must be larger than of order $mm^{-1}$, unless these modes
have suppressed coupling to matter compared with  the  zero-mode. We will discuss these issues in a concrete example in
Section 6.
\begin{table}[h]
\begin{center}
\begin{tabular}{|c|c|c|c|c|c|c|}
\hline
\multicolumn{1}{|c|}{ }
&
\multicolumn{1}{|c|}{$y\in (0,\infty)$}
&
\multicolumn{5}{|c|}{$y\in (0,y_0)$}\\
\hline
\multicolumn{1}{|c|}{$B(y_0)$}&
\multicolumn{1}{|c|}{$-$}&
\multicolumn{3}{|c|}{$-\a\, log (y_0-y)$}&
\multicolumn{2}{|c|}{$finite + (y_0-y)^\b$}
\\
& &$0<\a<1/2$ &$1/2<\a<1$&$1<\a<3/2$&$0<\b<1$&$1<\b<2$\\
\hline
Spin-2& $-$&$\bigcirc$&$\bigcirc$&$\bigcirc$&$\bigcirc$&$\bigcirc$\\
\hline
Spin-1& $-$&$\bigcirc$&$-$&$-$&$\bigcirc$&$\bigcirc$\\
\hline
Spin-0& $-$&$-$&$-$&$-$&$-$&$\bigcirc$\\
\hline
$V_t(y_0)$&$-$&$-1 /(y_0-y)^2$&$-1/(y_0-y)^2$&$+1/ (y_0-y)^2$& $+\infty$ & $-\infty$ \\
\hline
$\psi_t(y)$&$ -$& $(y_0-y)^\a$& $(y_0-y)^{1-\a}$&$1/(y_0-y)^{\a-1}$& finite &finite\\
\hline
$R(y_0)$&$-$&$+\infty$&$+\infty$&$-\infty$& $-\infty$ & $+\infty$\\
\hline
\end{tabular}
\caption{4D massless spectrum as a function of the IR behavior of $B(y)$.
A line $-$ or a circle $\bigcirc$ indicate  absence or presence, respectively,  of the corresponding
normalizable zero-mode. For the parameter ranges not shown in the table there is no normalizable spin-2 zero-mode.
The last three lines indicate the behavior of the Shr\"odinger potential, the tensor mode wave-function and the scalar
 curvature near the end of the $y$-coordinate range.}
\end{center}
\end{table}

The end of space can be either a boundary at an arbitrary location, or a singularity.
In the latter case the position of the singularity is not arbitrary, but it is determined by the background 
field equations and  boundary  conditions on the scale factor and dilaton. Depending on the type of singularity,
we can have different kinds of normalizable massless modes, as can be seen in table 1. Notice that when a mode
 does not appear it is always because it is a non-normalizable 4D state, therefore its would-be 4D kinetic term is
  infinite and the mode decouples from the low energy sector. This is unlike
  what was happening in \cite{gherghetta}: there, there was a massless
 scalar mode  which had \emph{zero}, rather than infinite, kinetic term,
and did not appear at all in the action at quadratic level. This mode is expected to become
strongly coupled when interactions are included.
In our case instead there is no danger of having strongly coupled light
scalar modes in the interacting theory.

Some of the cases analyzed here  cannot  be taken too seriously:
take for example the logarithmic  case, with $\a>1/2$:
  if we look at the amplitude of the metric perturbation, we find that the total metric is
\beq
g_{\mu\nu} = a^2 \left(\eta_{\mu\nu} + \left[a^{-3/2} \psi_t(y)\right] h_{\mu\nu}(x) \right) \sim   a^2 \left(\eta_{\mu\nu} +
 (y_0-y)^{1-2\a} h_{\mu\nu}(x) \right),
\eeq
 so for $\a>1/2$  the fluctuations blow up w.r.t. the background metric  near the singularity and the
 perturbation theory expansion  breaks down. This does not happen for $0<\a<1/2$ as well as for the finite
 scale factor cases, in which the perturbation always stays small.
\section{Dealing with the Singularity}

In the previous section we encountered different kinds
of singularities,  with different behavior of the
curvature scalars (see Table 1).
 It would be desirable to study which of these
singularities are intrinsically
pathological, and which  can be cured with  a physically
 meaningful resolution.

As sometimes happen  in AdS/CFT, the presence of a singularity
 in the bulk signals the occurrence of some nontrivial
infrared dynamics of the dual theory \cite{KT}, which once taken
 properly into account should produce a regular space-time
\cite{KS}. This is expected to be generic in confining theories
\cite{GK}.  Equivalently, one can hope that the IR singularity
will be regularized in a full string theory realization.
In any case one should check that the results one obtains do
not depend too much on the details of the regularization.
On the field theory side this means that what happens to the
IR of the theory can be  reconstructed from the detailed
 dynamics of the microscopic degrees of freedom\footnote{This
is analogous to the idea that given the fundamental QCD
 Lagrangian at some UV scale, one could in principle  calculate
the entire IR spectrum and dynamics without any
  further assumptions about the low energy physics.}, encoded
 holographically in the bulk action and  boundary
  conditions close to the $AdS$ boundary, where the bulk
 theory is under control.

In our case unfortunately the situation is not so simple.
In some of the cases we consider  the gravitational fluctuations
stay finite all the way  to the singularity, and bulk
 perturbation theory does not breakdown. However, we will
 see that the bulk dynamics plus boundary conditions are
 not enough to determine the low energy spectrum, but
  some additional  information about the IR must be given as
 input, in the form of specific boundary conditions at the singularity.
   This makes the question of regularization subtle, since
only for specific regularizations the results we
    achieve will be recovered. Therefore,  our initial
program of obtaining a massless spin-2 particle as part
    of the low energy spectrum of a well defined UV theory
is only partially successful.

\subsection{Boundary Conditions}

The simplest example we found that allows spin-2  zero-modes
 is  the one in which  the space has an
 $AdS$-like boundary  at  $y\sim 0$ and $B(y)$ diverges
 logarithmically as $B\sim -\a \log(y-y_0)$
 close to some point $y=y_0$: for $0<\a<3/2 $, equation
(\ref{sc2}) admits normalizable solutions with $m^2=0$.
This is not the end of the story, however: consider again
 the eigenvalue equation  for generic $m^2$:
\beq\label{eigenvalue}
-\psi'' + \left( B'^2 -B''\right)\psi = m^2 \psi.
\eeq
Close to the singularity at $y=y_0$, we can ignore the
$m^2$-term and the equation simplifies to
\beq
 \psi'' + {\a^2 -\a \over (y_0 -y)^2}\psi \sim  0,
\eeq
which is solved by simple power-laws:
\beq\label{asympt}
\psi \sim  c_1  (y_0 -y)^\a + c_2  (y_0 -y)^{1-\a}, \qquad y\sim y_0 .
\eeq
The crucial observation is that , in  the range of $\a$ in
 which we found zero-mode solutions, $0<\a<3/2$,
\emph{both terms in eq.} (\ref{asympt}) \emph{are square-integrable}.
 This means that the spectrum is not yet
 determined by the data given so far, but we need to further specify
 boundary conditions at the singularity.
 Unless we do so, we would find that eq. (\ref{eigenvalue}) admits
solutions for $m^2$ equal to \emph{any}
 complex  number\footnote{More technically,  we are dealing with
a second order differential operator which
is  not self-adjoint, since we are taking its domain
 to be too large.  To make it self-adjoint we need to restrict
 it on functions that obey specific boundary conditions
 at $y\sim y_0$. This is necessary  since it is self-adjointness
 that guarantees discreteness of the spectrum and
  orthogonality of the eigenfunctions associated to different
 eigenvalues. See e.g. \cite{richt} for technical details},
  and that the corresponding eigenfunctions are not orthogonal
 when inserted back  into the action. To make the problem
   well posed, i.e. to ensure that the operator we are considering
 has a discrete, non-negative spectrum and  orthogonal
    eigenfunctions, we have to specify boundary conditions at the
 singularity. It is this choice of boundary conditions
     that ultimately determines the spectrum\footnote{These extra
boundary conditions are not necessary at $y\sim 0$:
      there the two solutions behave as $y^{5/2}$ and $y^{-3/2}$,
only the first one of which is normalizable: the
       normalizability requirement is enough to restrict the
 choice to the solution that vanishes at $y=0$.}
In the case at hand, the boundary value of the eigenfunctions
is not well defined (it is either zero or infinity,
unless $\a=1$), but boundary conditions can be replaced by
conditions on the asymptotic behavior around $y=y_0$, i.e.
by fixing the coefficients $c_1$ and $c_2$ that appear in eq.
 (\ref{asympt}) \cite{richt}.

From the discussion above, it is clear that a zero
 eigenvalue exists as part of the spectrum only if the right
boundary conditions are imposed: these are the ones
 such that the zero-mode solution (\ref{zero2-2}) satisfies them.

Next, we analyze what happens if we start from the
boundary conditions for which the spectrum contains a spin-2 zero mode,
 and we slightly perturb them. Consider the asymptotic
behavior around $y=y_0$ given by (\ref{asympt}). For a
  specific value $c_1/c_2 = r_0$ the Schr\"odinger equation
(\ref{eigenvalue}) admits a zero-energy solution.
   Now we study the same problem with a slightly modified
boundary condition:
\beq\label{pert}
{c_1 \over c_2} = r_0 + \e, \qquad \e \ll 1.
\eeq
As in usual perturbation theory, we look for solutions of
(\ref{eigenvalue}) of the form:
\beq
\psi(y) = \psi^{(0)} + \psi^{(1)}
\eeq
where  $\psi^{(0)}$ is the zero-mode, satisfying
unperturbed asymptotic conditions (in practice
 $\psi^{(0)}\equiv\psi^{IR}$), and $\psi^{(1)}$ is taken to be small.
Then, the equation for $\psi_1(y)$ is, to lowest order:
\beq
-\psi^{(1)''} +  \left( B'^2 -B''\right)\psi^{(1)}  = m^2 \psi^{(0)}.
\eeq
This is solved, formally, by
\beq\label{psi1}
\psi^{(1)}(y) = m^2 \int d y'\, G_0 (y,y') \psi^{(0)}(y'),
\eeq
where $ G_0 (y,y') $ is  the Green's function of the one-dimensional
 problem with zero-eigenvalue:
\beq\label{green}
G_0 (y,y') =  {1\over 2} \left[ \psi^{IR}(y)\psi^{UV}(y')\theta(y'-y)+ \psi^{IR}(y)\psi^{UV}(y')\theta(y-y')\right],
\eeq
and  $\psi^{UV}$ and $\psi^{IR}$ are the independent solutions of
 the homogeneous equation, and are
explicitly given by (\ref{zero2})\footnote{One can verify
immediately that (\ref{green}) satisfies
\beq
-\de_{y'}^2 G_0 (y,y')  +  \left( B'^2 -B''\right)(y') G_0 (y,y') = \delta(y-y'), \nn
\eeq
using the Wronskian relation $\psi^{IR'}\psi^{UV}-\psi^{IR}\psi^{UV'} =1$.
 The ordering is fixed
requiring that the solution (\ref{psi1}) is normalizable in the UV.
There is an ambiguity from the
fact that we can always add to $\psi_1$ a multiple of $\psi_0$ and
still get a solution, since the
latter is a zero-mode, but this shift is proportional to the
 unperturbed solution and it is therefore
trivial. It can be fixed requiring e.g. that the unperturbed solution
is normalized to one}.

Close to the singularity,  $y\sim y_0$ only the first part of the
Green's function contributes,
 so  the  perturbed solution behaves  asymptotically as:
\bea
\psi &\sim& \psi^{IR}(y) + m^2 \psi^{UV}(y) \int d y'\, \left|\psi^{IR} (y')\right|^2 \nn\\ &=&
 c_1^{(0)} (y_0-y)^\a +  c_2^{(0)} (y_0-y)^{1-\a} + m^2  (y_0-y)^\a
\eea
where  $c_1^{(0)}$ and $c_2^{(0)}$ are the expansion
 coefficients of the unperturbed solution, satisfying
 by assumption $c_1^{(0)}/ c_2^{(0)}=r_0$, and we have
used the explicit form of $\psi^{UV}(y)$ in the
 last equality. Now we impose the perturbed boundary conditions, eq. (\ref{pert}) and find that the mass
 eigenvalue is given, to leading order, by:
\beq
m^2 = \e \,c_2^{(0)}.
\eeq

This result shows that a small deformation away of the
 boundary conditions lifts the zero-mode to
 a light massive mode. Nevertheless the choice of boundary
 condition is \emph{not} continuous in the
 $\e\to 0$ limit: we have seen in Section 4 that the massless
 and massive bulk fluctuations contain
 different number of degrees of freedom, at least for in the
 interesting cases, when no scalar modes
 are normalizable: for example, in the logarithmic case, there
are at most four physical degrees of
 freedom in the massless sector, while as we have seen it takes
 a total of six to have a light massive
  multiplet. This means that  the boundary conditions that give
 rise to a massless graviton are special,
   and once they are tuned that way the situation is completely
stable, since  it corresponds to a
   reduced number of physical modes.

One last comment. If we want to fix   boundary conditions using
 the asymptotic behavior in (\ref{asympt}),
 for any  given scale factor $a(y)$ it is not clear how to
specify the ``zero-mode-friendly'' boundary conditions
  \emph{a priori}, and in a purely local way, since it appears that
 the subleading behavior of the function in eq.
  (\ref{zero2-2}) (corresponding to  the coefficient $c_1$, if $\a>1/2$)
 carries information about the behavior of the
   $a(y)$ away from the singularity. An alternative, and equivalent
way to impose  boundary conditions is to
   relate the leading UV and IR behaviors: the condition\beq
\lim_{y\to 0} \left[e^{-2B}\left(e^B \psi\right)'\right] =
 \lim_{y\to y_0}\left[ e^{-2B}\left(e^B \psi\right)'\right],
\eeq
or equivalently
\beq
\lim_{y\to 0} \left[a^3 h'\right]= \lim_{y\to y_0}\left[ a^3 h'\right],
\eeq
is sufficient to fix the spectrum, and it is obeyed by
the zero-mode wave-function.
 It has the advantage that it does not require detailed
 knowledge of the scale factor in the bulk.

\section{A Concrete Example}

In this section we consider an  explicit example which gives rise to
IR-emergent massless gravitons. The fifth dimension  terminates
at a singularity at  $y=y_0$, and  we derive explicitly the
special boundary conditions  to be  imposed at that point in such a way that
a spin-2 massless mode is in the 4D spectrum.

Consider the simplest example
when $B(y)$ has  $AdS$-like asymptotics  in the UV and  logarithmically divergent in the IR:
\beq
B(y)= {3\over 2}\log k y - \a \log (1-y/y_0), \qquad 0<\a<3/2.
\eeq
The background  metric reads:
\beq
ds^2 = {(1-y/y_0)^{{4\over 3}\a}\over (k y)^2 }\left (dy^2 + \eta_{\mu\nu} d x^\mu d x^\nu\right).
\eeq
The space-time is asymptotically $AdS_5$ near $y=0$, with $AdS$ curvature scale given by $k$.
We saw in the previous sections that, provided suitable boundary conditions
are imposed at the $y=y_0$ singularity, the massless spectrum in 4D contains  just
one spin-2 mode (if $1/2<\a<3/2$ ) or one spin-2 and one  spin-1 mode
(if $0<\a<1/2$). The massless graviton is given by:
\beq\label{profile}
h_{\mu\nu}(y,x) = a^{-3/2}(y)\psi_t(y) h_{\mu\nu}^{(4)}(x) ,
\eeq
where the profile wave-function is
\beq
\psi_t(y) = e^{-B(y)}\int_0^y e^{2B(y)} = y_0 (k y_0)^{3/2} F(y/y_0),
\eeq
and $F(z)$ is a dimensionless function of $z=y/y_0$:
\beq\label{explicit_wf}
F(z) = {(1-z)^\a \over z^{3/2}}\int_0^z dz'\, {z'^3\over (1-z')^{2\a}}.
\eeq
It behaves asymptotically as:
\bea
&& F_\a(z) \sim z^{5/2} \qquad \qquad \qquad \qquad \qquad \,z \sim 0, \label{smallz} \\
&& F_\a(z) \sim c_1 (1-z)^\a + c_2(1-z)^{1-\a} \quad z\sim 1.
\eea

Near the singularity, at $z=1$, we  have:
\beq
h  = a^{-3/2} \psi_t \sim  \left[c_1 + O(1-z) \right] +
  (1 -z )^{1-2\a}\left[c_2 + O(1-z) \right]
\eeq
We see that if $\a>1/2$ the metric perturbation diverges
close to the singularity, so strictly
 speaking a perturbative treatment is not fully justified.
 In the following we assume $\a<1/2$: in
  this case $h(y)$ goes to a constant $c_1$ and the profile $\psi$ vanishes close to $y_0$.

Expanding eq. (\ref{explicit_wf}) near $y/y_0=1$ we can get
the ratio $c_1/c_2$, which a posteriori
determines what boundary condition we have to impose so that
 the zero-mode is in the spectrum (recall
from the previous  section that the boundary condition can be set
 precisely by specifying this ratio.) We
find that the boundary conditions that one needs to impose to
keep the zero-mode in the spectrum are:
\beq
c_1 = 3(1-\a)(2-\a)(2-3\a)(1-2\a) c_2
\eeq

Now suppose that ``we'' (i.e.  the visible SM sector)
are confined on a 3-brane at a fixed position $y=y_B$
in the geometry just considered. We are going to work in
the probe-brane approximation, ignoring the backreaction
of the SM brane on the geometry. Then, 4D gravity between
 brane-localized sources will be mediated by the
graviton zero-mode. Higher KK  modes of mass $m$ will
 contribute only below distances $1/m$. This mass scale
is set by the parameter $y_0$, i.e. generically we will have
 $m^{-1}\sim y_0$, therefore if $y_0$ is small
enough these modes will not contribute to the gravitational
attraction at large distances. Below we will make
this argument more precise.

\begin{center}
\begin{figure}[h]
\epsfxsize=5in 
\epsffile{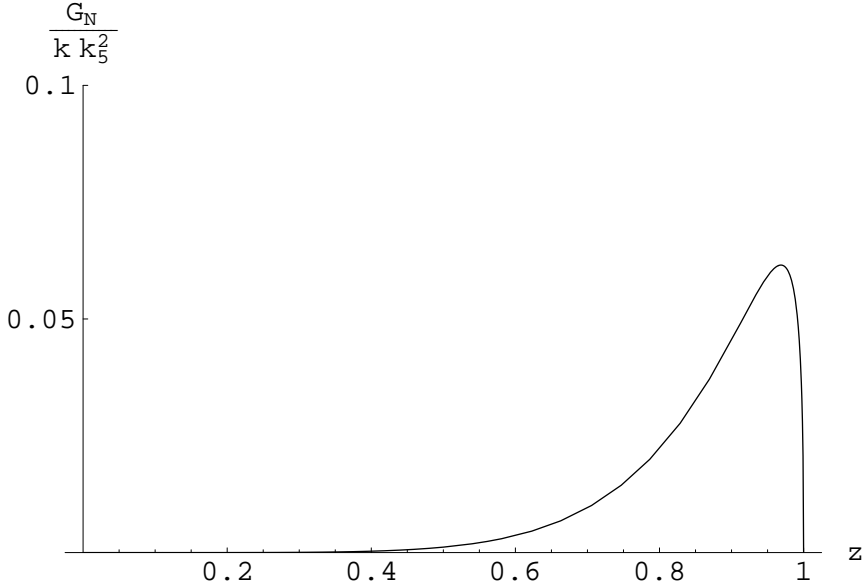}
\caption{Four Dimensional Newton's Constant as a function of
the position of the visible brane, for $\a=0.25$}
\label{Newton}
\end{figure}
\end{center}

Next, we determine the effective 4D gravitational coupling constant
 $G_N$ on the brane. The action for brane-matter is
\beq
S_{brane} = \int_{y=y_B}{\mathcal L}_{brane}.
\eeq
  The 4D energy-momentum  $T_{\mu\nu}$ couples canonically
to the induced  metric $\hat{g}_{\mu\nu}$, and is defined by
\beq
T_{\mu\nu} = -{1\over \sqrt{-\hat{g}}}{\delta S \over \delta \hat{g}^{\mu\nu}}.
\eeq
To linear order in the fluctuations,
$\hat{g}_{\mu\nu}= a^2(y_B)(\eta_{\mu\nu} + h_{\mu\nu})$, so the
 brane source term, expanded to linear order in $h_{\mu\nu}$, has the form:
\beq
S_{brane}=\int {\delta  {\mathcal L}_{brane}\over
 \delta \hat{g}^{\mu\nu}}\delta  \hat{g}^{\mu\nu} =
\int \sqrt{-\hat{g}} T_{\mu\nu}  \left(a^2 h^{\mu\nu}\right).
\eeq
 The physical coordinates measured on the brane,
in which the induced  metric is just $\eta_{\mu\nu}$,
 differ from  the bulk coordinate $x^\mu$ by a rescaling,
$x^\mu \to a^{-2}(y_B) x^\mu$. After this rescaling,
the effective 4D action including the brane source  is:
\bea
S && = {1\over 2 k_5^2} \int d y d^4 x\, {a^3(y)\over a^2(y_B)}
 \left(\de h_{\mu\nu}(y,x)\right)^2
 + \int_{y=y_B} d^4 x\,  h_{\mu\nu}(y_B,x) T^{\mu\nu}(x) \nn\\
&& \!\!\!\!\!\!\!\!\!\!=  {1\over 2 k_5^2 \, a^2(y_B)} \int d y \,
 \psi_t^2(y)\int  d^4 x\,
\left(\de h_{\mu\nu}^{(4)}\right)^2 + a^{-3/2}(y_B) \psi_t(y_B)
\int d^4 x\,h_{\mu\nu}^{(4)} T^{\mu\nu}(x), \label{effective}
\eea
where in the last line we have used eq. (\ref{profile}).
Normalizing canonically the 4D graviton $h_{\mu\nu}^{(4)}(x)$ we can read off from eq. (\ref{effective})
the effective 4D Newton's constant:
\beq\label{gnewton}
(8\pi G_N)^{1/2} = k_5  {a^{-1/2}(y_B) \psi(y_B) \over
\left(\int_0^{y_0}\psi^2(y) \right)^{1/2}} =
k_5 \sqrt{k} {z_B^{1/2}\over (1-z_B)^{\a/3}}{F(z_B)\over {\mathcal N}},
\eeq
where $z_B=y_B/y_0$ and ${\mathcal N}= (\int_0^1 F^2)^{1/2}$.
An example of the behavior of $G_N$ as a
function of $z_B$ is displayed in figure \ref{Newton}.
For small $z_B$ (i.e. if the brane is close to the $AdS$ boundary),
$G_N$ vanishes as $z_B^3$, so gravity
 effectively decouples close to the boundary. In the dual interpretation, this corresponds
to the fact that the graviton, being a composite, ceases to exists at high energy, and its
interactions become soft in the UV.

Using the small $z$ behavior (\ref{smallz}) in eq. (\ref{gnewton}), we can see
that  the effective  4D Plank scale, $M_p = 1/\sqrt{(8\pi G_N)}$, felt on a brane close to the
$AdS$ boundary is:
\beq\label{mplanck}
M_p^2 \sim {M_5^3 \over k} \left({y_0 \over y_B}\right)^6.
\eeq
This expression does not depend on this specific example,
 as the UV behavior of the zero-mode wave-function is
independent of what happens in the IR. So if we assume we
live on a brane close to the UV boundary, the low-energy
 phenomenology of all the models we are considering has
the same universal behavior, and will start to differ at
 energies when the lowest KK states become relevant. This will happen
roughly at energies of order $1/y_0$, so the value of $y_0$ is constrained
by small scale gravity measurement to be less than say $1 mm$.
 This can be easily
 achieved by taking, e.g. (if one is thinking about
 having the weak scale as the fundamental Planck scale in the bulk)
  $M_5 \sim k \sim 1/y_B \sim 1 \,TeV$ and $y_0^{-1} \sim 2\, GeV$.
\section{Conclusion and Possible  Generalizations}

We have investigated the presence of a massless four-dimensional graviton from a 5D covariant  theory.
This may also be interpreted as a massless spin-2
glueball in the spirit of bulk/boundary correspondence.
Our (general) result is that in an asymptotically $AdS_5$  space-time with 4D Lorentz covariance,
one cannot have normalizable massless 4D spin-2
 modes unless the space-time terminates at either a brane
or at a singularity. We worked in the context of Einstein-Dilaton gravity, but  our  result is completely general:
the linearized equation for the tensor polarization
 is sensitive only to the background metric,
and  one can obtain \emph{any} physically reasonable
 (i.e. not  containing ghosts)
 background solution by choosing  the dilaton potential
 appropriately. Therefore  our result for
 the spin-2 modes does not depend at all on the field content
 of the 5D theory. One could go even
 further, and allow for space-times that do not satisfy the
 positivity condition of Sec. 4.1,
  as will be the case if we include scalar fields with ``wrong'' sign kinetic term, or the
  equivalent of a ``phantom'' perfect fluid matter in the bulk.
This would give rise to more
   general possibilities for the background solution,
 which are not  necessarily pathological:
    for example, if the kinetic term of some scalar
 is controlled by the vev of another field,
     and has the wrong sign only in some limited region of space-time.

On the positive side, we did find many examples  of
solutions with normalizable massless spin-2 modes.
 They contain either an IR  boundary, or a IR singularity.
 In both cases one is faced with the
  question of how to determine boundary conditions.
The usual choices (e.g. Neumann b.c.) do not give a massless graviton.
Despite appearances, the singular case seems more appealing than the ``hard wall'' boundary.
 From the $AdS$/CFT perspective, it provides a dynamical way
to end space-time, which can be completely
  encoded in the UV data of the theory. Also, in many cases
the singularity renders most or all the unwanted
  ``extra'' polarizations, that are usually present in a
higher-dimensional theory, non-normalizable.

The presence of the singularity raises various
questions, and one starts
wondering about possible resolutions that preserve
the interesting boundary conditions. It is
 unlikely that the usual methods - terminating the
space-time with an Euclidean black hole horizon, or going to higher
 dimension and adding vanishing cycles - will work,
since they usually lead to Neumann b.c, for which there
 is no graviton zero-mode\footnote{We can use
for example the argument in  \cite{gubser}:
 the zero-mode wave equation can be written
as $(a^3 h')'=0$, so if $a^3 h'$ is non-zero somewhere,
  it will be non-zero everywhere. Now,  the
normalizable zero-mode, eq.  (\ref{zero2-2}),  has $a^3 h'\neq 0$ at the $AdS$
   boundary, so it  cannot satisfy Neumann b.c. $h'=0$
at any point with a  finite value of the scale factor.}.

As the theory living in the bulk is supposed to
be coming from a  string theory, the  singularity should be resolved in the full theory.
One can ask whether the kinds of solutions we studied in
this work admit a realization within string theory.  There are several examples
of string theoretical constructions in ten dimensions that give rise to
asymptotically $AdS_5$,   singular backgrounds
when they are reduced to D=5. Some of the resulting five-dimensional geometries fit in the
class that, according to the analysis of this paper, potentially admit massless spin-2
modes as the only low-energy states in D=4. To quote one example, one of the solutions of 5D
supergravity considered in \cite{sfetsos} (called ``model B'' in that paper)  has an
 asymptotically AdS region\footnote{The particular setup  considered in \cite{sfetsos}
 excised that region  as in RS, and did not extend all the way to the AdS boundary,
 in order to render the constant graviton wave-function normalizable.},
 and a singularity at finite $y$ close to  which the scale factor behaves like in
our eq. (\ref{logwarp}), with $\a=1/2$:  it therefore
 belongs in the class that, according to our analysis (see Section \ref{summary}),
admit a massless graviton localized in the IR. This geometry is a solution
of a particular supergravity  theory in 5D which comes from a truncation
of D=10 supergravity on a background that arises as the near-horizon geometry
of D3-branes  uniformly distributed over an $S^3$ \cite{freedman2}.
 Given that there
are string theoretical configurations  that give rise to similar singularities
to the ones we considered, it would be interesting to study in these examples
what kind of boundary conditions are obtained once
 the full string theoretical resolution of the singularity is reduced
to the 5D effective theory.

Our result, that no massless four-dimensional gravitons arise in generic
warped extra-dimensional models with $AdS_5$ asymptotics,  reflects  the
Weinberg-Witten no-go theorem
\cite{ww} in the dual field theory. This forbids  the occurrence of massless spin-2 states
in a 4D theory  with a Lorentz-covariant stress tensor.
This theorem is based on several assumptions,  and relaxing one or more
of these assumptions is expected to give different results.
It would be interesting to understand in which way the cases when we
have indeed found a massless spin-2 state circumvent the theorem. One
of the assumptions of the theorem, that gravity couples universally to
the 4D field theory stress tensor, is certainly not satisfied
in our models, since according to the standard AdS/CFT paradigm,  the
``fundamental'' 4D field theory stress tensor is sourced by 
 the \emph{non-normalizable}
 graviton zero-mode. There are other assumptions we can relax to
find more generic scenarios that could avoid the difficulties we found
in this work, and may allow to consider non-singular space-times.
 We can envision at least three possible directions, all of which can be easily
 addressed in our formalism. They  all  correspond to relaxing some assumption
of the Weinberg-Witten theorem.

\vspace{0.3cm}
\emph{Give up $m^2 = 0$}
\vspace{0.3cm}

If we look for massive, rather than massless,  spin-2
 normalizable modes, the arguments
in section 4 do not apply: these modes are of course
present generically in our model no
 matter what the IR behavior of the scale factor, so we
can deal with regular space-times that
 do not end. It is easy to obtain a purely discrete
spectrum, for example choosing $B(y) \sim y^2$
 in the IR, which is clearly allowed by the positivity
condition. This gives a quadratic potential
 for the tensor  fluctuations, and the requirement of
normalizability at $y=0$ and $y=+\infty$
 are enough to determine the spectrum without need of
any extra boundary condition.
 The spectrum will look like the one needed for
``linear confinement'' in $AdS$/QCD \cite{Karch:2006pv}.
 Of course, if we allow a mass for the graviton, we face
a different class of phenomenological problems.
 The  minimal requirement for the model to have some chance
 is that  the graviton mass is tiny and there is large mass gap
 between the ground state (massive  graviton) and the first
excited KK states (in order for the theory to not become strongly coupled at low energies).
 This does not
 seem impossible to achieve, as we are not aware of any
general argument that forbids it, but it
 is technically difficult.
 The main obstacle is the positivity requirement (\ref{positive2}): if that
 is lifted, and we allow ``wrong sign'' kinetic
 terms for bulk scalars even in some small region,  then most
likely it will not be difficult  getting
 such a spectrum, but in  this case one can  presumably
adjust things so that the ground state is massless after all.

Assuming one can generate the  correct hierarchy between
the ground state and the excited states,  one
would have a very interesting model, which at low energy
reduces to massive gravity,
 and whose UV completion and covariantizations are
 controllable. One can then explore
 unambiguously the questions such as strong coupling
regime and deviation from GR in
  the field of large sources \cite{georgi,Porrati:2004mz,dvali}

  Another possibility connected to this line of reasoning, is generating a tiny graviton mass
 via "boundary" interactions with another universe, as
  was recently advocated in \cite{clo,aha}.
  Indeed, in such a context the stress tensor is not conserved due to $1/N^2$ effects,
   and this generates a mass for the graviton in a controllable fashion.

\vspace{0.3cm}
\emph{Give up Normalizability}
\vspace{0.3cm}

We can also ask whether we can obtain long-lived 4D spin-2
massless  \emph{resonances},
rather than one-particle states. Lifting the normalizability
 requirement allows once more to
 include regular space-times, with infinite range of $y$.
 One way this could
 work is having a space that looks like one of our ``good''
singular examples almost all the way to
  the singularity, and patch it continuously  with some
regular space  that extends all the way to
  $y=+\infty$. This will turn  the infinite potential
 wall (or well), felt at the singularity
   by the tensor wave-function, into a  wall of finite
height (or well of finite depth).
    Then the tensor Schr\"odinger potential appearing in eq. (\ref{sc2})  would have
to approach zero at $y=\infty$  as we
    need a continuum of light states for the resonance
to decay. In this case we need the
     resonance to be very narrow, and to have significantly
enhanced coupling to matter w.r.t.
      the other continuum modes.

\vspace{0.3cm}
\emph{Give up Lorentz Invariance}
\vspace{0.3cm}

We have chosen to consider a 5D space-time with
exact 4D Lorentz symmetry. If we give up this
requirement the Schr\"odinger potential felt by the
tensor fluctuations changes,
 as seen for example in \cite{karchrandall} in the
 context of a pure $AdS$ bulk. To start with,
  one could limit the analysis to maximally symmetric
4D space-times, and consider 5D metrics that
 are conformally $\Real \times (A)dS_4$.
\vspace{0.3cm}

 There are other phenomenological issues that one must clarify  in our model. The first concerns
 the nonlinear interaction of the graviton zero-mode.
It is not obvious a priori that, in the models we consider, these
reproduce the graviton self-interactions of four-dimensional
General Relativity, and this requirement is an additional constraint
that realistic models should satisfy. For example, in this model,
the coefficient $\kappa_g$ of the cubic self-interaction of zero-mode graviton,
and the coupling constant governing the interaction of gravity with 4D matter
(the effective Newton's constant $G_N$), are two independent parameters: the first
depends only on the bulk geometry, whereas
the second depends also on the position of the brane. A phenomenologically
viable model can result only if these two parameters are related in the same
way as in ordinary 4D Einstein's gravity, where $G_N = \kappa_g^2$.
This relation can always be recovered by placing
 the observable brane at an appropriate position, and it would be interesting
to see if there is some model in which this is achieved naturally.
One could worry about higher order interactions, but the one
arising from the cubic gravitational self-interaction is really the only
constraint, as this is  the only nonlinear term of the Einstein's equation
that is actually tested by direct gravity measurements.

Related to the  question of higher order interactions is
the fact that, above energies when it is possible to probe the
extra dimension, the equivalence principle will generically be violated,
because coupling of the zero-mode to the visible sector is
universal only insofar the latter is confined at a fixed value
of the $y$ coordinate. We will leave  these issues  for discussion in future
work.

\vskip 1cm
\begin{flushleft}
{\large \bf Acknowledgments}
\end{flushleft}
\addcontentsline{toc}{section}{Acknowledgments}
It is a pleasure to thank A. Bucher, G. Dvali, T. Gherghetta, S. Giddings, U. Gursoy, M. Peloso, M. Porrati, A. Veinshtein for discussions.

The work was partially supported by ANR grant NT05-1-41861,
INTAS grant, 03-51-6346,
RTN contracts MRTN-CT-2004-005104 and MRTN-CT-2004-503369,
CNRS PICS 2530 and 3059 and by a European Excellence Grant,
MEXT-CT-2003-509661. F. N. is supported by a  European
Commission Marie Curie Intra European Fellowship, contract
 MEIF-CT-2006-039369

\newpage
\appendix
\addcontentsline{toc}{section}{Appendices}
\section*{APPENDIX}

\section{Quadratic Action and Linearized Perturbations}
\renewcommand{\theequation}{A.\arabic{equation}}
We define the fluctuations around the background (\ref{b0}) as:
\bea\label{fluc0app}
 ds^2 && = a^2(y)\left(\eta_{AB} + h_{AB}\right) \nn\\
&& =a^2(y)\left[\left(1+2\phi\right)dy^2 + 2A_{\mu}dydx^{\mu}+
 \left(\eta_{\mu\nu} + h_{\mu\nu}\right)dx^\mu dx^\nu\right], \\
\Phi && = \Phi_0(y) + \chi \nn
\eea
where $\phi$, $A_{\mu}$, $h_{\mu\nu}$ and $\chi$ are functions of $y,x^{\mu}$.

To write the action for these fluctuations at quadratic order it is convenient to
  use the conformal properties of the curvature scalar: for a metric $g_{AB}= e^{-A}\tilde{g}_{AB}$ we have:
\beq
\sqrt{-g}R = e^{-(D/2-1)A} \sqrt{-\tilde{g}}\left[\tilde{R} + (D-1)\tilde{g}^{AB}
 \tilde{\nabla}_A \tilde{\nabla}_B A - {(D-2)(D-1) \over 4} \tilde{g}^{AB}\tilde{\nabla}_A A \tilde{\nabla}_B A\right].
\eeq
Specializing to $D=5$ ,  defining $e^{-A}$ = $a^2$ and neglecting a total derivative we can rewrite the action, eq. (\ref{action}), in the form
\beq\label{confaction}
S = {1\over 2k_5^2}\int d^4x dy\, e^{-3/2\,A} \sqrt{-\tilde{g}}\left[  \tilde{R} + 3
 \tilde{g}^{AB} \de_A A \de_B A - \tilde{g}^{AB} \de_A \Phi \de_B \Phi - e^{-A} V(\Phi)\right]
\eeq
Now we expand this action to quadratic order in the fluctuations (\ref{fluc0}), using
$\tilde{g}_{AB}= \eta_{AB} + h_{AB}$ and  expanding $\sqrt{-\tilde{g}}$ and $\tilde{R}$ around a flat background:
\bea
\tilde{g}_{AB} & =& \eta_{AB} + h_{AB}, \quad \tilde{g}^{AB}= \eta^{AB} - h^{AB} + h^{AC}h_C^B; \nn\\
 \sqrt{-\tilde{g}} & =& 1 + {1\over 2} h - {1\over 4}\left(h_{AB}h^{AB} - {1\over 2} h^2\right), \quad  h \equiv h^A_A;
 \nn\\
 \tilde{R}_{AB}& = & \de^C \de_{\left(A\right.} h_{\left.B\right)C} - {1\over2}\de^C\de_C h_{AB}
 -{1\over2} \de_A\de_B h \nn\\
& +& {1\over2} h^{CD}\de_A \de_B h_{CD} - h^{CD}\de_C\de_{\left(A\right.} h_{\left.B\right)D} +
{1\over 4} (\de_A h_{CD}) (\de_B h^{CD}) + {1\over2}  (\de^D h_B^C) (\de_D h_{AC}) \nn\\
& -& {1\over2}  (\de^D h_B^C) (\de_C h_{AD})- (\de_D h^{CD})(\de_{\left(A\right.} h_{\left.B\right)C} )
 + {1\over2}\de_D(h^{DC}\de_C h_{AB}) \nn \\
& -&  {1\over 4}(\de^C h)(\de_C h_{AB}) +{1\over2}(\de^C h) (\de_{\left(A\right.} h_{\left.B\right)C} ).
\eea
Indexes are contracted with $\eta_{AB}$; the last expression can be found e.g. in  \cite{wald}. Plugging into
(\ref{confaction}), using the background equations and  some integration by parts, one gets:
\bea
S &=&  {1\over 2k_5^2}\int d^4x dy\, a^3(y) \nn\\
&& \left[-{1\over4} (\de^C h_{AB})(\de_C h^{AB}) + {1\over2 }(\de^B h_{AB})(\de_C h^{AC}) -{1\over2}
(\de_A h)(\de_C h^{AC}) + {1\over4} (\de_A h)(\de^A h)\right. \nn\\ && +\left. \Phi_0' h'\chi + 2
\Phi_0'h^{Ay}\de_A \chi - \de_A\chi \de^A \chi -{1\over2} a^2 \de^2_{\Phi}V \, \chi^2\right]\nn\\
 &&- \left(a^3(y)\right)' \left[\de^Ch_{Ay} h^A_C - (h^{Ay})'h_{Ay} + {1\over2}\de^A h h_{Ay}\right].
\eea
Using the decomposition (\ref{fluc0app}) this becomes:
\bea\label{action2app}
S^{(2)} &=&  {1\over 2k_5^2}\int d^4x dy\, a^3(y)  \Bigg[ L_{ein}^{(2)}  - {1\over4} h'_{\rho\sigma}h'^{\rho\sigma} +
{1\over4} (h')^2 -{1\over4}F_{\mu\nu}F^{\mu\nu} \nn\\
& -&  \de_{\mu}\chi \de^{\mu}\chi -\chi'^2- {1\over2}a^2 \de^2_{\Phi}V \, \chi^2 -
\de^{\mu}\phi\left(\de^{\nu}h_{\mu\nu} - \de_\mu h\right) \nn\\
&+& 2 \Phi_0' \phi' \chi + \Phi_0' h' \chi + 4 \Phi_0' \phi \chi' + 2 \Phi_0' A^{\mu}\de_{\mu}\chi \Bigg] \nn\\
&-& \left(a^3 A^{\mu}\right)' \left[\de_{\mu}h - \de^{\nu}h_{\mu\nu}\right] \nn\\
& +& \left(a^3 \right)' \left[ -2 A_{\mu}\de^\mu\phi - 2 \phi\phi' - \phi h'\right],
\eea
where from now on $h\equiv h_\mu^\mu$, $F_{\mu\nu} = \de_\mu A_\nu - \de_\nu A_\mu$,  and
\beq\label{einsteinapp}
 L_{ein}^{(2)} = -{1\over 4}\de^\mu h_{\rho\sigma}\de_\mu h^{\rho\sigma} + {1\over2}
 \de^\mu h_{\rho\mu}\de_\nu h^{\rho\nu}- {1\over2} \de^\mu h \de_\rho h^{\rho\mu} +
{1\over 4} \de^\mu h \de_\mu h,
\eeq
is the  quadratic part of the 4D Einstein-Hilbert Lagrangian.

The field equations derived from this action are:
\bea
(\mu\nu) &&  h_{\mu\nu}'' + 3{a'\over a} h_{\mu\nu}' + \Box h_{\mu\nu} -2 \de^{\rho}
\de_{\left(\mu\right.}h_{\left.\nu\right)\rho} +\de_\mu\de_\nu h + 2\de_\nu\de_\mu \phi -2
 a^{-3}\left(a^3 \de_{\left(\mu\right.} A_{\left.\nu\right)}\right)' \nn\\
&& + \eta_{\mu\nu}\Bigg[-h'' - 3{a'\over a}h' - \Box h -2 \Box \phi + 6{a'\over a}\phi' +
6\left({a''\over a}+2 \left({a'\over a}\right)^2\right)\phi  \nn\\
&& -2 a^{-3}\left(a^3 \Phi_0' \chi\right)' + 2 a^{-3}\left(a^3 \de_{\mu} A^{\mu}\right)' +
\de^\rho \de^\sigma h_{\rho\sigma}\Bigg] =0 ;\label{munu}\\
(\mu y) && \Box A_\mu - \de^\nu h_{\nu\mu}'  \nn\\
&& +\de_\mu\left[ - 6{a'\over a}\phi + 2\Phi_0' \chi + h' - \de^\nu A_\nu\right] =0 ;\label{muy}\\
(yy) && -\Box h + \de^{\mu}\de^\nu h_{\mu\nu} - 3{a'\over a}h' + 6 \left({a''\over a}+2
 \left({a'\over a}\right)^2\right)\phi \nn\\
&& + 6 {a'\over a} \de^\nu A_\nu - 2 a^{-3}\left(a^3 \Phi_0' \chi\right)' + 4 \Phi_0' \chi' = 0 ;\label{yy}\\
(Dil) && \chi'' + 3 {a'\over a}\chi' + \Box \chi - {1\over2}a^2 \de^2_{\Phi}V \chi \nn\\
&&  -2 a^{-3} \left(a^3 \Phi_0' \phi\right)' +\Phi_0'\phi' + {1\over2}\Phi_0'h'   -
 \Phi_0' \de^\mu A_\mu=0. \label{dilaton}
\eea

\subsection{Massive Modes}
This case was analyzed in detail in \cite{peloso}. If $ k^2\neq 0$ one can further
 decompose $A_\mu$ and $h_{\mu\nu}$ in irreducible representations of the 4D Lorentz group in the following way:
\bea
&&A_{\mu} = \de_\mu W + A^{T}_\mu, \quad  \de^\mu A^{T}_\mu=0 \label{split1} \\
&&h_{\mu\nu} = 2\eta_{\mu\nu}\psi + 2 \de_\mu\de_\nu E + 2 \de_{\left(\mu\right.}
V^{T}_{\left.\nu\right)} + h^{TT}_{\mu\nu},\label{split2}
\eea
 with  $\de^\mu V^{T}_\mu= \de^\mu h^{TT}_{\mu\nu} = h^{TT\mu}_{\mu}=0$; the fields
$W$,$\psi$ and $E$ are additional Lorentz-scalars, to be added to the already defined scalar fluctuations
$\phi$ and $\chi$ (see eq. (\ref{fluc0app}); the fields  $ A^{T}_\mu$ and $ V^{T}_\mu$ are Lorentz-vectors.
 If $ k^2\neq 0$ this decomposition is unique, i.e. one can invert the above relations and
write all these fields as functions of $A_\mu$ and $h_{\mu\nu}$. The gauge transformation
properties of these fields can be read-off from (\ref{gauge1}-\ref{gauge2}): under
\beq
(\delta x^\mu, \delta y) = (\xi^{\mu}, \xi^5) \equiv (\xi^{T\mu} + \de^\mu \xi, \xi^5), \quad \de_\mu\xi^{T\mu}=0,
\eeq
the fields defined in (\ref{split1}-\ref{split2}) change according to:
\beq\label{gauge5}
\delta \psi = -{a'\over a}\xi^5, \quad \delta E =-\xi, \quad \delta V^T_\mu = \xi^T_\mu,
\quad   \delta  h^{TT}_{\mu\nu}=0, \quad \delta W = -\xi' -{a'\over a}\xi^5, \quad  \delta A^T_\mu = \xi^{T'}_\mu.
\eeq

The field equations (\ref{munu}-\ref{dilaton}) can split in separate  equations involving
the scalar, vector and TT-tensor only:
\begin{itemize}
\item tensor modes:
\beq\label{tensorgeneric}
  h_{\mu\nu}^{TT''} + 3{a'\over a} h_{\mu\nu}^{TT'} + \Box h_{\mu\nu}^{TT}=0;
\eeq
\item vector modes:
\beq\label{vectorgeneric}
 \Box (A_\mu^T- V^{T'}_\mu)=0, \quad \left(a^3 (A_\mu^T- V^{T'}_\mu)\right)'=0;
\eeq
\item scalar modes:
\bea
\!\!\!\!0&=&\psi'' + 3 {a'\over a} \psi' - {a'\over a}\phi' - \left({a''\over a}+2
 \left({a'\over a}\right)^2\right)\phi + {1\over 3} a^{-3}\left(a^3 \Phi_0' \chi\right)',\label{s1} \\
\!\!\!\!0&=&  \phi + 2 \psi - (W-E')' -3 {a'\over a}(W-E'),\label{s2} \\
\!\!\!\!0&=& \psi' - {a'\over a}\phi + {1\over3}\Phi_0' \chi ,\label{s3}\\
\!\!\!\!0&=&\Box \psi  +4 {a'\over a}\psi'- \left({a''\over a}+2 \left({a'\over a}\right)^2\right)\phi- {a'\over a} \Box(W-E')\nn\\
&& - {1\over3} a^{-3}\left(a^3 \Phi_0' \chi\right)' - {2\over3} \Phi_0' \chi',\label{s4}\\
\!\!\!\!0&=&\chi'' + 3 {a'\over a}\chi' + \Box \chi - {1\over2}a^2 \de^2_{\Phi}V
\chi -2 a^{-3} \left(a^3 \Phi_0' \phi\right)' \nn\\
&& +\Phi_0'\phi' + 4\Phi_0'\psi'   - \Phi_0' \Box(W-E').\label{s5}
\eea
\end{itemize}
Notice that only the combinations $A_\mu^T- V^{T'}_\mu$ and $W-E'$ appear in the field equations.
 Also, notice that there are no massive vector modes.

Let us consider the scalar modes. Not all the equations are independent: indeed eq. (\ref{s1})
 is a consequence of (\ref{s3}). One can solve the system eliminating $\phi$ using (\ref{s3})
and $\Box(W-E')$ using (\ref{s4}) in terms of $\psi$ and $\chi$. Substituting their expressions in
(\ref{s5}) only on the following linear combination of $\psi$ and $\chi$ appears:
\beq
\zeta = \psi - {\chi\over z},\quad \quad  z\equiv {a \Phi_0' \over a'}, \label{zeta}
\eeq
and  equation (\ref{s5}) reduces to \cite{peloso}:
\beq
\zeta'' + \left(3 {a'\over a} + 2 {z'\over z}\right) \zeta'+ \Box \zeta = 0, \label{zetaeq}\\
\eeq
The fact that only a linear combination of $\psi$ and $\chi$ is determined by the field equations
 is a consequence of the gauge freedom $\delta y = \xi^5(x,y)$: it can be used to set either $\psi$ or
 $\chi$ to an arbitrary function, as one can see from (\ref{gauge4}) and the first of (\ref{gauge5}),
and only the gauge invariant combination $\zeta$ has a physical meaning.

\subsection{Massless Modes}\label{masslessmodes}

When looking for solutions of the form (\ref{separation}) with  $k^2=0$, the decompositions
(\ref{split1}, \ref{split2}) become ambiguous:
since  all the terms on the l.h.s.
(except the one proportional to $\eta_{\mu\nu}$ are transverse and traceless.
 Equivalently, it is impossible to solve for $E$, $W$, $V_\mu$ and $\psi$
in terms of $h_{\mu\nu}$ and $A_\mu$, since the resulting expressions would contain inverse powers of $\Box$, e.g. $W = \Box^{-1} \de^\mu A_\mu$.
What we need is a decomposition of vectors and
 tensor along $k_\mu$ and three other independent 4-vectors. If $k_\mu$ is timelike, as in the
previous subsection, one can take as a basis $k_\mu$ and three spacelike vectors orthogonal to it,
resulting in the decompositions  (\ref{split1}, \ref{split2}).  If $k_\mu$ is null, on the other
hand, the basis can be taken as $(k_\mu^+,k_\mu^-,k_\mu^i)$, $i=1,2$, defined by the relations:
\beq
k_\mu^+\equiv k_\mu, \quad   k_\mu^- k^{+\mu} =1,\quad  k_\mu^- k^{-\mu} = 0, \quad k_\mu^i k^{+\mu} =
   k_\mu^i k^{-\mu}=0, \quad  k_\mu^i k^{j\mu}=\delta^{ij}\eeq
One  has the completeness relation:
\beq\label{completeness}
\eta_{\mu\nu} = k_\mu^+k_\nu^- +  k_\mu^-k_\nu^+ +  k_\mu^i k^{j}_\nu \delta_{ij}.
\eeq
The expansions of a vector and a symmetric tensor along this basis take the form\bea
&&A_\mu = k^+_\mu A^+ +   k^-_\mu A^- + \sum_{i=1,2} k_\mu^iA^i ,\\\label{Amu}
&&h_{\mu\nu}=  k^+_\mu k^+_\nu h^{++} +  2k^+_{\left(\mu\right.} k^-_{\left.\nu\right)} h^{+-} +
 k^-_\mu k^-_\nu h^{--}  \nn\\
&& + \sum_{i=1,2} 2k^+_{\left(\mu\right.} k^i_{\left.\nu\right)}h^{+i} +  \sum_{i=1,2} 2k^-_{\left(\mu\right.}
 k^i_{\left.\nu\right)}h^{-i}  +  \sum_{i,j=1,2} k^i_{\left(\mu\right.} k^j_{\left.\nu\right)}h^{ij}. \label{hmunu}
\eea
Under an infinitesimal diffeomorphism  with parameters $(\xi^\mu ,\xi^5)$ we have:
\beq
\delta h_{\mu\nu} = -\de_\mu\xi_\nu - \de_\mu\xi_\nu - 2 \eta_{\mu\nu}{a'\over a}\xi_5.
\eeq
It is useful to split $h_{\mu\nu}$ in two parts, as
\beq
h_{\mu\nu} = \tilde{h}_{\mu\nu} + 2\eta_{\mu\nu}\psi,
\eeq
such that $ \tilde{h}_{\mu\nu}$ and $\psi$  transform separately under 4D and 5D diffeomorphisms, that is:
\beq
\delta \tilde{h}_{\mu\nu} =  -\de_\mu \xi_\nu - \de_\nu \xi_\mu, \quad \delta \psi= - {a'\over a} \xi^5.
\eeq
From eq. (\ref{completeness}),  $\eta_{\mu\nu}$ 
contains only $+-$ and $ij$ components , so the only 
components that can appear in the definition of $\psi$ are 
$h^{+-}$ and $h^{ij}$. These are also the only components in eq. 
(\ref{hmunu}) that transform under a $\xi_5$-diffeomorphism:
\beq
 \delta h^{+-} = -2 {a'\over a}\xi_5, \qquad \delta h^{ij} = -2 {a'\over a}\delta^{ij}\xi_5.
\eeq
For $\psi$ to transform correctly, it must be of the form
$\psi =  h^i_i/4  + A (h^{+-} - h^i_i/2)$,
for some real number $A$, since the second term is invariant;
 since $h^{+-}$ transforms non-trivially under $\xi_\mu$-diffeomorphisms,
we must take $A=0$. So we have:
\beq
\psi = {1\over 4} h^i_i, \qquad  \tilde{h}_{\mu\nu} = h_{\mu\nu}- {1\over 2} \eta_{\mu\nu} h^i_i.
\eeq
Notice that, in a decomposition like that of eq. (\ref{hmunu}),
 the $ij$ part  of $\tilde{h}_{\mu\nu}$ is traceless.

In order to solve the field equations we first fix the gauge for $A_\mu$ and $\tilde{h}_{\mu\nu}$.
Under a gauge transformation, we have:
\beq
\delta (\de^\mu \tilde{h}_{\mu\nu}) = - \Box \xi_\nu - \de_\nu \de^\rho \xi_\rho, \quad \delta \tilde{h}_{\mu}^{\mu}
 = -2  \de_\nu \de^\rho \xi_\rho, \quad \delta (\de^\mu A_\mu) = -\de^\mu \xi_\mu' - \Box \xi_5
\eeq
therefore
\beq
\delta \left(\de^\mu \tilde{h}_{\mu\nu}-{1\over2}\de_\nu\tilde{h}_{\mu}^{\mu} \right) = - \Box \xi_\nu,
\eeq
and we can set,  by an appropriate choice of $\xi_\nu$ and $\xi_5$:
\bea
 &&\de^\mu \tilde{h}_{\mu\nu}-{1\over2}\de_\nu\tilde{h}_{\mu}^{\mu}  =0. \label{ddgauge}\\
&& \de^\mu A_\mu =0. \label{lorentzgauge}
\eea
In terms of the decompositions (\ref{Amu}, \ref{hmunu}),
this gauge  corresponds to  $A^-= h^{--} =  h^{-i} =0$.

The gauge choice (\ref{ddgauge}, \ref{lorentzgauge}) is left intact by gauge parameters that satisfy
$\Box \xi_\mu = \Box \xi_5 =0$. This implies that $\xi_\mu$ and $\de_\mu\xi_5$ can also be decomposed in a ``light-cone''   basis as:
\beq
\xi_\mu = k_\mu^+ \xi^+ +  k_\mu^- \xi^- +  k_\mu^i \xi^i, \quad \de_\mu\xi_5 = k_\mu^+ \xi_5.
\eeq
We can  use this residual gauge freedom, to set $\tilde{h}^{++}= \tilde{h}^{-+}=   \tilde{h}^{+i}=0$ and
 $A^+=0$. This additional choice implies $\de^\mu\tilde{h}_{\mu\nu}=\tilde{h}_{\mu}^{\mu}=0$

To summarize, we can fix  the following gauge:
\bea
&&A_\mu = A^T_\mu \qquad \de^\mu A_\mu = 0\\
&&h_{\mu\nu} = h^{TT}_{\mu\nu} + 2 \eta_{\mu\nu}\psi, \qquad \de^\mu h_{\mu\nu}=2\de_\nu\psi,\;
 h_{\mu}^{\mu} = 8\psi,\label{gaugechoice}\\
&&  h^{TT}_{\mu\nu} = \sum_{i,j=1,2}k^i_\mu k^j_\nu \left(h_{ij}-{1\over2}\delta_{ij}h_l^l\right),
\quad  A^T_\mu =  \sum_{i=1,2}k^i_\mu A^i.
\eea
Notice that  $ h^{TT}_{\mu\nu}$ and $A^T_\mu$ have  components only along $i,j$, i.e.
 the directions orthogonal to the \emph{spatial} momentum, so they have only two independent components each.
 This gauge choice corresponds to keeping only the physical helicity states of the massless spin-2 and spin-1 fields.

Using these gauge conditions, and the fact that we are considering zero-modes of the 4D wave operator,
the field equations (\ref{munu}-\ref{dilaton}) become:
 \bea
(\mu\nu) &&  h_{\mu\nu}^{TT''} + 3{a'\over a} h_{\mu\nu}^{TT'} + \de_\mu\de_\nu(4\psi + 2\phi) - 2
\left(a^3 \de_{\left(\mu\right.}A^T_{\left.\nu\right)}\right)'\nn\\
&& + \eta_{\mu\nu}\Bigg[-6\psi'' - 18{a'\over a}\psi' + 6{a'\over a}\phi' + 6\left[{a''\over a}+2
 \left({a'\over a}\right)^2\right]\phi  \nn \\
&&  -2 a^{-3}\left(a^3 \Phi_0' \chi\right)'\Bigg] =0 \label{munu0}\\
(\mu y)&& \de_\mu\left[ - 6{a'\over a}\phi + 2\Phi_0' \chi + 6\psi'\right] =0 \label{muy0}\\
(yy) && - 12{a'\over a}\psi' + 3 \left[{a''\over a}+2 \left({a'\over a}\right)^2\right]\phi -
 a^{-3}\left(a^3 \Phi_0' \chi\right)' + 2 \Phi_0' \chi' = 0 \label{yy0}\\
(Dil) && \chi'' + 3 {a'\over a}\chi'  - {1\over2}a^2 \de^2_{\Phi}V \chi \nn\\
&&  -2 a^{-3} \left(a^3 \Phi_0' \phi\right)' +\Phi_0'\phi' + 4\Phi_0'\psi'=0. \label{dilaton0}
\eea
Taking the trace of eq. (\ref{munu0}) we see that the two  lines of that equation have to vanish
 individually. The vanishing of the first line can be written in the ``light-cone'' basis as:
\beq
k^i_\mu k^j_\nu \left[ h_{ij}'' +  3{a'\over a} h_{ij}'\right] - 2k^i_{\left(\mu\right.}
k^+_{\left.\nu\right)} \left[(a^3 A_i)'\right]  + k^+_\mu k^+_\nu\left[4\psi + 2\phi\right]=0.
\eeq
The three terms in the equation above are independent, so  have to vanish separately. This yields
 equations for the transverse tensor and vector zero-modes, and an equation for the longitudinal
component of the vector:
\bea
&&  h_{\mu\nu}^{TT''} +  3{a'\over a} h_{\mu\nu}^{T'} = 0 \label{tensorzero}\\
&&  (a^3 A_\mu^T)' = 0 \label{vectorzero}\\
&&  2\psi + \phi =0  \label{scalarzero}
\eea
The last equation eliminates one out of the three  scalars ($\chi$ , $\psi$,  $\phi$), so we are
left with a total of six propagating fields, which is the correct number of degrees of freedom in our setup.

Eqs. (\ref{tensorzero}) and (\ref{vectorzero}) give immediately the tensor and vector profiles in
the $y$-direction. To get the scalar modes wave-functions we need a bit more work.
First, consider eq. (\ref{muy0}): if the 4-momentum is not strictly zero, the quantity in square brackets
 has to vanish. Combined with eq. (\ref{scalarzero}), this gives
\beq \label{scalar1}
 \psi' =-2 {a' \over a} \psi- {1\over 3} \Phi_0' \chi. 
\eeq
Substituting  into eq. (\ref{yy0}) and using repeatedly eq. (\ref{scalarzero}), we arrive at:
\beq \label{scalar2}
\psi = -{1\over 2} {\chi \over z} - {1\over 2}\left({a\over a'} {\chi \over z}\right)'
\eeq
As a consistency check, one can easily show that the two equations above imply the last
remaining scalar equation, (\ref{dilaton0}).

After some manipulations, the two equations (\ref{scalar1}, \ref{scalar2}) can be shown
to be equivalent to the system:
\bea
&& \zeta_1' = 0, \qquad  \left({a^4\over a'} \zeta_2 \right)' = -2 a^3 \zeta_1 \label{system} \\
&& \zeta_1 \equiv \psi -{\chi\over z}; \qquad \zeta_2 \equiv {\chi\over z} \label{zetas}
\eea
where  $z(y)$ is given by (\ref{zed}).

\end{document}